\documentclass[article,nojss]{jss}
\usepackage{thumbpdf,lmodern} 
\graphicspath{{Figures/}}

\usepackage{amsmath,amssymb,enumitem,subcaption}
\usepackage[figuresright]{rotating}

\newcommand{\ones}{\mathbf 1}
\newcommand{\reals}{{\mbox{\bf R}}}
\newcommand{\integers}{{\mbox{\bf Z}}}
\newcommand{\symm}{{\mbox{\bf S}}}  

\newcommand{\cvxr}{\pkg{CVXR}}

\title{\cvxr{}: An \proglang{R} Package for Disciplined Convex Optimization}

\author{Anqi Fu\\Stanford University \And Balasubramanian Narasimhan\\Stanford University \And Stephen Boyd\\Stanford University}

\Plainauthor{Anqi Fu, Balasubramanian Narasimhan, Stephen Boyd} 
\Plaintitle{CVXR: An R Package for Disciplined Convex Optimization} 

\Abstract{ \cvxr{} is an \proglang{R} package that provides an
  object-oriented modeling language for convex optimization, similar
  to \pkg{CVX}, \pkg{CVXPY}, \pkg{YALMIP}, and \pkg{Convex.jl}. It
  allows the user to formulate convex optimization problems in a
  natural mathematical syntax rather than the restrictive form
  required by most solvers. The user specifies an objective and set of
  constraints by combining constants, variables, and parameters using
  a library of functions with known mathematical properties. \cvxr{}
  then applies signed disciplined convex programming (DCP) to verify
  the problem's convexity. Once verified, the problem is converted
  into standard conic form using graph implementations and passed to a
  cone solver such as \pkg{ECOS} or \pkg{SCS}. We demonstrate
  \cvxr{}'s modeling framework with several applications.  }


\Plainkeywords{convex optimization, disciplined convex optimization,
  optimization, regression, penalized regression, isotonic regression,
  R package CVXR}

\Address{
  Anqi Fu\\
  Department of Electrical Engineering\\
  David Packard Building\\
  350 Jane Stanford Way\\
  Stanford, CA 94305\\
  E-mail: \email{anqif@stanford.edu}\\
  URL: \url{https://web.stanford.edu/~anqif/}\\

  Balasubramanian Narasimhan\\
  Department of Biomedical Data Sciences, and\\
  Department of Statistics\\
  Stanford University\\
  390 Jane Stanford Way\\
  Stanford, CA 94305\\
  E-mail: \email{naras@stanford.edu}\\
  URL: \url{https://statistics.stanford.edu/~naras/}\\

  Stephen Boyd\\
  Department of Electrical Engineering\\
  David Packard Building\\
  350 Jane Stanford Way\\
  Stanford, CA 94305\\
  E-mail: \email{boyd@stanford.edu}\\
  URL: \url{https://web.stanford.edu/~boyd/}\\
}

\begin{document}

\section{Introduction}
\label{intro}
Optimization plays an important role in fitting many statistical
models. Some examples include least squares, ridge and lasso
regression, isotonic regression, Huber regression, support vector
machines, and sparse inverse covariance
estimation. \mbox{\citet{koenker:mizera:2014}} discuss the role of
convex optimization in statistics and provide a survey of packages for
solving such problems in \proglang{R} \mbox{\citep{r:2018}}. Our
package, \cvxr{} \citep{CVXR}, solves a broad class of convex
optimization problems, which includes those noted above as well as
many other models and methods in statistics.
Similar systems already exist, such as \pkg{CVX} \citep{cvx}
and \pkg{YALMIP} \citep{YALMIP} in \proglang{MATLAB}
\citep{MATLABR2019a}, \pkg{CVXPY} {\citep{CVXPY}} in
\proglang{Python} \citep{python}, and \pkg{Convex.jl}
{\citep{cvxjl}} in \proglang{Julia} \citep{julia}.  \cvxr{}
brings these capabilities to \proglang{R}, providing a domain-specific
language (DSL) that allows users to easily formulate and solve new
problems for which custom code does not exist.  As an illustration,
suppose we are given $X \in \reals^{m \times n}$ and $y \in \reals^m$,
and we want to solve the ordinary least squares (OLS) problem
\[
  \begin{array}{ll}
    \underset{\beta}{\mbox{minimize}} & \|y - X\beta\|_2^2
  \end{array}
\]
with optimization variable $\beta \in \reals^n$. This problem has a
well-known analytical solution, which can be determined
using \code{lm} in the default \pkg{stats} package. In \cvxr{}, we can
solve for $\beta$ using the code
\begin{CodeChunk}
\begin{CodeInput}
R> beta <- Variable(n)
R> obj <- sum((y - X %*% beta)^2)
R> prob <- Problem(Minimize(obj))
R> result <- solve(prob)
\end{CodeInput}
\end{CodeChunk}
The first line declares our variable, the second line forms our
objective function, the third line defines the optimization problem,
and the last line solves this problem by converting it into a
second-order cone program and sending it to one of \cvxr{}'s
solvers. The results are retrieved with
\begin{CodeChunk}
\begin{CodeInput}
R> result$value             # Optimal objective
R> result$getValue(beta)    # Optimal variables
R> result$solve_time        # Solver runtime
\end{CodeInput}
\end{CodeChunk}
This code runs slower and requires additional set-up at the
beginning. So far, it does not look like an improvement on \code{stats::lm}. However,
suppose we add a constraint to our problem:
\[
	\begin{array}{ll}\underset{\beta}{\mbox{minimize}} & \|y - X\beta\|_2^2 \\
	\mbox{subject to} & \beta_j \leq \beta_{j+1}, \quad j = 1,\ldots,n-1.
	\end{array}
\]
This is a special case of isotonic regression. Now, we can no longer
use \code{stats::lm} for the optimization. We would need to find
another \proglang{R} package tailored to this type of problem {such as
  \pkg{nnls} \citep{nnls}} or write our own custom solver. With
\cvxr{} though, we need only add the constraint as a second argument
to the problem:
\begin{CodeChunk}
\begin{CodeInput}
R> prob <- Problem(Minimize(obj), list(diff(beta) >= 0))
\end{CodeInput}
\end{CodeChunk}
Our new problem definition includes the coefficient constraint, and a
call to \code{solve} will produce its
solution. In addition to the usual results, we can get the dual
variables with
\begin{CodeChunk}
\begin{CodeInput}
R> result$getDualValue(constraints(prob)[[1]])
\end{CodeInput}
\end{CodeChunk}
This example demonstrates \cvxr{}'s chief advantage:
flexibility. Users can quickly modify and re-solve a problem, making
our package ideal for prototyping new statistical methods. Its syntax
is simple and mathematically intuitive. Furthermore, \cvxr{} combines
seamlessly with native \proglang{R} code as well as several popular
packages, allowing it to be incorporated easily into a larger
analytical framework. The user can, for instance, apply resampling
techniques like the bootstrap to estimate variability, as we show in
Section~\ref{ex:nonpar}.

DSLs for convex optimization are already widespread on other
application platforms. In \proglang{R}, users have access to the
packages listed in the CRAN Task View for {\em Optimization and
  Mathematical Programming} {\mbox{\citep{cran:opt}}}. Packages like
\pkg{optimx} \citep{optimx} and \pkg{nloptr} \citep{nloptr} {provide
  access to} a variety of general algorithms, {which can handle
  nonlinear and certain classes of nonconvex problems. \cvxr{}, on the
  other hand, offers a language to express convex optimization
  problems using \proglang{R} syntax, along with a tool for analyzing
  and restructuring them for the solver best suited to their
  type}. \pkg{ROI} \citep{ROI} is perhaps the package closest to ours
in spirit. It offers an object-oriented framework for defining
optimization problems, but still requires users to explicitly identify
the type of every objective and constraint, whereas \cvxr{} manages
this process automatically.

In the next section, we provide a brief mathematical overview of
convex optimization. Interested readers can find a full treatment in
\citet{BoydVandenberghe:2004}. Then we give a series of examples
ranging from basic regression models to semidefinite programming,
which demonstrate the simplicity of problem construction in
\cvxr{}. Finally, we describe the implementation details before
concluding. Our package and the example code for this paper are
available on the Comprehensive \proglang{R} Archive Network (CRAN) at
\url{https://CRAN.R-project.org/package=CVXR} and the official \cvxr{}
site at \url{https://cvxr.rbind.io}.

\section{Disciplined convex optimization}
\label{convex}
The general convex optimization problem is of the form
\[
\begin{array}{ll}
	\underset{v}{\mbox{minimize}} & f_0(v)\\
	\mbox{subject to} & f_i(v) \leq 0, \quad i=1,\ldots,M\\
	& Av=b,
\end{array}
\]
where $v \in \reals^n$ is our variable of interest, and
$A \in \reals^{m \times n}$ and $b \in \reals^n$ are constants describing our linear equality constraints. The objective and
inequality constraint functions $f_0,\ldots,f_M$ are convex, i.e., they
are functions $f_i: \reals^n \rightarrow \reals$ that satisfy
\[
	f_i(\theta u + (1-\theta)v) \leq \theta f_i(u) + (1-\theta)f_i(v)
\]
for all $u,v \in \reals^n$ and $\theta \in [0,1]$. This class of
problems arises in a variety of fields, including machine learning and
statistics.

A number of efficient algorithms exist for solving convex problems
\citep{Wright:1997, ADMM, AndersenDahl:2011, SkajaaYe:2015}. However,
it is unnecessary for the \cvxr{} user to know the operational details
of these algorithms. \cvxr{} provides a DSL that allows the user to
specify the problem in a natural mathematical syntax. This
specification is automatically converted into the standard form
ingested by a generic convex solver. See Section~\ref{implement} for more
on this process.

In general, it can be difficult to determine whether an optimization
problem is convex. We follow an approach called disciplined convex
programming \citep[DCP;][]{GrantBoydYe:2006} to define problems using a
library of basic functions (atoms), whose properties like curvature,
monotonicity, and sign are known. Adhering to the DCP rule,
\begin{center}
  $f(g_1,\ldots,g_k)$ is convex if $f$ is convex and for each $i = 1,\ldots,k$, either
  \begin{itemize}[leftmargin=30mm]
  \item $g_i$ is affine,
  \item $g_i$ is convex and $f$ is increasing in argument $i$, or
  \item $g_i$ is concave and $f$ is decreasing in argument $i$,
  \end{itemize}
\end{center}
we combine these atoms such that the resulting problem is
convex by construction. Users will need to become familiar with this rule if they wish to define complex problems.

The library of available atoms is provided in the documentation. It
covers an extensive array of functions, enabling any user to model and
solve a wide variety of sophisticated optimization problems. In the
next section, we provide sample code for just a few of these problems,
many of which are cumbersome to
prototype or solve with other \proglang{R} packages.

\section{Examples}
\label{examples}
In the following examples, we are given a dataset $(x_i,y_i)$ for
$i = 1,\ldots,m$, where $x_i \in \reals^n$ and $y_i \in \reals$. We
represent these observations in matrix form as
$X \in \reals^{m \times n}$ with stacked rows $x_i^\top$ and
$y \in \reals^m$. Generally, we assume that $m > n$.

\subsection{Regression}
\label{ex:regression}

\subsubsection{Robust (Huber) regression}
\label{ex:huber}
In Section~\ref{intro}, we saw an example of OLS in \cvxr{}. While least
squares is a popular regression model, one of its flaws is its high
sensitivity to outliers. A single outlier that falls outside the tails
of the normal distribution can drastically alter the resulting
coefficients, skewing the fit on the other data points. For a more
robust model, we can fit a Huber regression \citep{Huber:1964} instead
by solving
\[
	\begin{array}{ll} \underset{\beta}{\mbox{minimize}} & \sum_{i=1}^m \phi(y_i - x_i^{\top}\beta) \end{array}
\]
for variable $\beta \in \reals^n$, where the loss is the Huber
function with threshold $M > 0$,
\[
	\phi(u) = 
	\begin{cases}
		\frac{1}{2}u^2 & \mbox{if } |u| \leq M \\
		M|u| - \frac{1}{2}M^2 & \mbox{if } |u| > M.
	\end{cases}
\]
This function is identical to the least squares penalty for small
residuals, but on large residuals, its penalty is lower and increases
linearly rather than quadratically. It is thus more forgiving of
outliers.

In \cvxr{}, the code for this problem is
\begin{CodeChunk}
\begin{CodeInput}
R> beta <- Variable(n)
R> obj <- sum(huber(y - X %*% beta, M))
R> prob <- Problem(Minimize(obj))
R> result <- solve(prob)
\end{CodeInput}
\end{CodeChunk}
Note the similarity to the OLS code. As before, the first line instantiates the $n$-dimensional optimization variable, and the second line defines the objective function by combining this variable with our data using \cvxr{}'s library of atoms. The only difference this time is we call the \code{huber} atom on the residuals with threshold \code{M}, which we assume has been set to a positive scalar constant. Our package provides many such atoms to simplify problem
definition for the user.

\subsubsection{Quantile regression}
\label{ex:quantile}
Another variation on least squares is quantile regression
\citep{quantile}. The loss is the tilted $l_1$ function,
\[
	\phi(u) = \tau\max(u,0) - (1-\tau)\max(-u,0) = \frac{1}{2}|u| + \left(\tau - \frac{1}{2}\right)u,
\]
where $\tau \in (0,1)$ specifies the quantile. The problem as before is to minimize the total residual loss. This model is commonly used in ecology, healthcare, and other fields where the mean alone is not enough to capture complex relationships between variables. \cvxr{} allows us to create a function to represent the loss and integrate it seamlessly into the problem definition, as illustrated below.
\begin{CodeChunk}
\begin{CodeInput}
R> quant_loss <- function(u, tau) 0.5 * abs(u) + (tau - 0.5) * u
R> obj <- sum(quant_loss(y - X %*% beta, t))
R> prob <- Problem(Minimize(obj))
R> result <- solve(prob)
\end{CodeInput}
\end{CodeChunk}
Here \code{t} is the user-defined quantile parameter. We do not need
to create a new `\code{Variable}' object, since we can reuse
\code{beta} from the previous example.

{By default, the \code{solve} method automatically selects the \cvxr{}
  solver most specialized to the given problem's type. This solver may
  be changed by passing in an additional \code{solver} argument. For
  instance, the following line fits our quantile regression with
  \pkg{SCS} \citep{SCS}.}
\begin{CodeChunk}
\begin{CodeInput}
R> result <- solve(prob, solver = "SCS")
\end{CodeInput}
\end{CodeChunk}
\subsubsection{Elastic net regularization}
\label{ex:elasticnet}
Often in applications, we encounter problems that require
regularization to prevent overfitting, introduce sparsity, facilitate
variable selection, or impose prior distributions on parameters. Two
of the most common regularization functions are the $l_1$-norm and
squared $l_2$-norm, combined in the elastic net regression model
\citep{elasticnet, glmnet},
\[
\begin{array}{ll} \underset{\beta}{\mbox{minimize}} & \frac{1}{2m}\|y - X\beta\|_2^2 + \lambda(\frac{1-\alpha}{2}\|\beta\|_2^2 + \alpha\|\beta\|_1). \end{array}
\]
Here $\lambda \geq 0$ is the overall regularization weight and
$\alpha \in [0,1]$ controls the relative $l_1$ versus squared $l_2$
penalty. Thus, this model encompasses both ridge ($\alpha = 0$) and
lasso ($\alpha = 1$) regression.

To solve this problem in \cvxr{}, we first define a function that
calculates the regularization term given the variable and penalty
weights.
\begin{CodeChunk}
\begin{CodeInput}
R> elastic_reg <- function(beta, lambda = 0, alpha = 0) {
+    ridge <- (1 - alpha) * sum(beta^2)
+    lasso <- alpha * p_norm(beta, 1)
+    lambda * (lasso + ridge)
+  }
\end{CodeInput}
\end{CodeChunk}
Then, we add it to the scaled least squares loss.
\begin{CodeChunk}
\begin{CodeInput}
R> loss <- sum((y - X %*% beta)^2)/(2 * m)
R> obj <- loss + elastic_reg(beta, lambda, alpha)
R> prob <- Problem(Minimize(obj))
R> result <- solve(prob)
\end{CodeInput}
\end{CodeChunk}
The advantage of this modular approach is that we can easily
incorporate elastic net regularization into other regression
models. For instance, if we wanted to run regularized Huber
regression, \cvxr{} allows us to reuse the above code with just a
single changed line,
\begin{CodeChunk}
\begin{CodeInput}
R> loss <- sum(huber(y - X %*% beta, M))
\end{CodeInput}
\end{CodeChunk}
\subsubsection{Logistic regression}
\label{ex:logistic}
Suppose now that $y_i \in \{0,1\}$ is a binary class indicator. One of
the most popular methods for binary classification is logistic
regression \citep{Cox:1958, Freedman:2009}. We model the
conditional response as $y|x \sim \mbox{Bernoulli}(g_{\beta}(x))$,
where $g_{\beta}(x) = \frac{1}{1 + e^{-x^\top\beta}}$ is the logistic
function, and maximize the log-likelihood function, yielding the
optimization problem
\[
\begin{array}{ll} \underset{\beta}{\mbox{maximize}} & \sum_{i=1}^m \{ y_i\log(g_{\beta}(x_i)) + (1-y_i)\log(1 - g_{\beta}(x_i)) \}.
\end{array}
\]

\cvxr{} provides the \code{logistic} atom as a shortcut for
$f(z) = \log(1 + e^z)$, so our problem is succinctly expressed as
\begin{CodeChunk}
\begin{CodeInput}
R> obj <- -sum(X[y == 0, ] %*% beta) - sum(logistic(-X %*% beta))
R> prob <- Problem(Maximize(obj))
R> result <- solve(prob)
\end{CodeInput}
\end{CodeChunk}

The user may be tempted to type \code{log(1 + exp(X \%*\% beta))} as
in conventional \proglang{R} syntax. However, this representation of $f(z)$
violates the DCP composition rule, so the \cvxr{} parser will reject
the problem even though the objective is convex. Users who wish to
employ a function that is convex, but not DCP compliant should check
the documentation for a custom atom or consider a different
formulation.

We can retrieve the optimal objective and variables just like in
OLS. More interestingly, we can evaluate various functions of these
variables as well by passing them directly into
\code{result\$getValue}. For instance, the log-odds are
\begin{CodeChunk}
\begin{CodeInput}
R> log_odds <- result$getValue(X %*% beta)
\end{CodeInput}
\end{CodeChunk}
This will coincide with the ratio we get from computing the probabilities directly:
\begin{CodeChunk}
\begin{CodeInput}
R> beta_res <- result$getValue(beta)
R> y_probs <- 1 / (1 + exp(-X %*% beta_res))
R> log(y_probs / (1 - y_probs))
\end{CodeInput}
\end{CodeChunk}
\begin{figure}[t!]
  \centering
  \includegraphics[width=\textwidth]{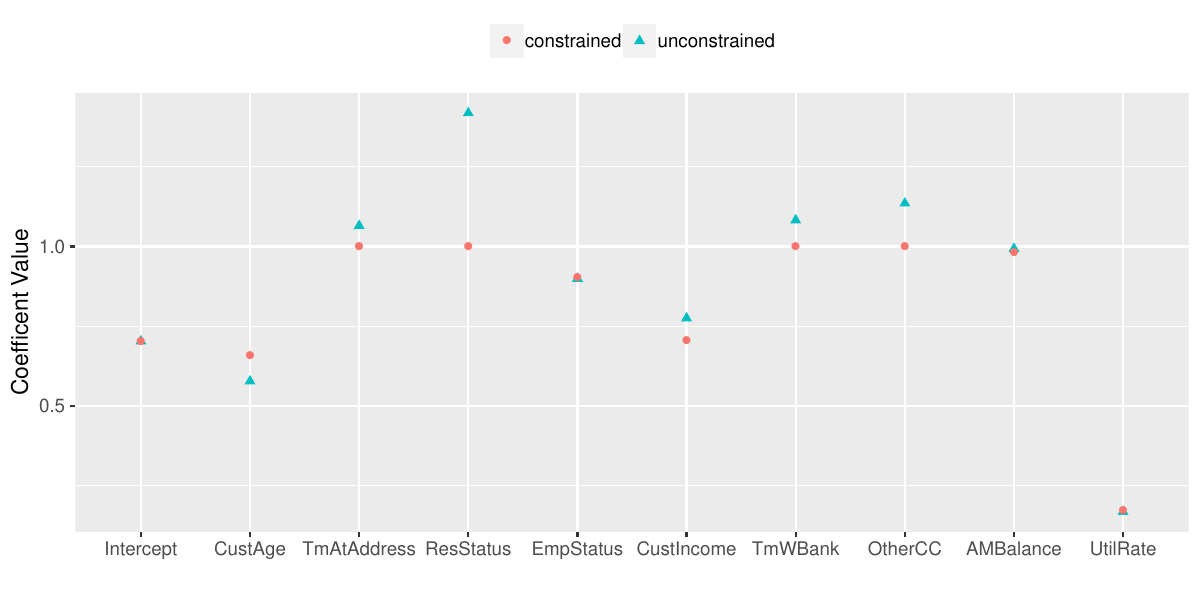}
  \caption{Logistic regression with constraints using data from
    \citet{credit}. The addition of
    constraint~\ref{eq:credit-corr-constraint} moves the coefficients
    for customer age and customer income closer to each other.}
  \label{fig:logistic}
\end{figure}

We illustrate with a logistic regression fit from a credit scoring
example \citep{credit}. The nine regression coefficients other than
the intercept are constrained to be in the unit interval.  To reflect
the correlation between two of the covariates, customer age ($x_2$)
and customer income ($x_6$), an additional constraint is placed on the
respective coefficients $\beta_2$ and $\beta_6$:
\begin{equation}
  \label{eq:credit-corr-constraint}
  |\beta_{2} - \beta_{6}| \leq 0.5.
\end{equation}
The code below demonstrates how the latter constraint can be
specified by seamlessly combining familiar \proglang{R} functions such
as \code{abs} with standard indexing constructs.
\begin{CodeChunk}
\begin{CodeInput}
R> constr <- list(beta[2:10] >= 0, beta[2:10] <= 1,
+    abs(beta[2] - beta[6]) <= 0.05)
R> prob <- Problem(Maximize(obj), constr)
R> result <- solve(prob)
R> beta_res_con <- result$getValue(beta)
\end{CodeInput}
\end{CodeChunk}
Figure~\ref{fig:logistic} compares the unconstrained and constrained
fits and shows that the addition of
constraint~\ref{eq:credit-corr-constraint} pulls the coefficient
estimates for customer age and customer income towards each other.

Many other classification methods {belong to} the convex
framework. For example, the support vector classifier is the solution
of a
$l_2$-norm minimization problem with linear constraints, which we have
already shown how to model. Support vector machines are a
straightforward extension. The multinomial distribution can be used to
predict multiple classes, and estimation via maximum likelihood
produces a convex problem. To each of these methods, we can easily add
new penalties, variables, and constraints in \cvxr{}, allowing us to
adapt to a specific dataset or environment.

\subsubsection{Sparse inverse covariance estimation}
\label{ex:invcov}
Assume we are given i.i.d.\ observations $x_i \sim N(0,\Sigma)$ for
$i = 1,\ldots,m$, and the covariance matrix $\Sigma \in \symm_+^n$,
the set of symmetric positive semidefinite matrices, has a sparse
inverse $S = \Sigma^{-1}$. Let
$Q = \frac{1}{m-1}\sum_{i=1}^m (x_i - \bar x)(x_i - \bar x)^\top$ be
our sample covariance. One way to estimate $\Sigma$ is to maximize the
log-likelihood with {an $l_1$-norm constraint \citep{YuanLin:2007,
    Banerjee:2008, spinvcov}}, which amounts to the optimization
problem
\[
\begin{array}{ll} \underset{S}{\mbox{maximize}} & \log\det(S) - \mbox{tr}(SQ) \\
\mbox{subject to} & S \in \symm_+^n, \quad \sum_{i=1}^n \sum_{j=1}^n |S_{ij}| \leq \alpha.
\end{array}
\]
The parameter $\alpha \geq 0$ controls the degree of sparsity. Our
problem is convex, so we can solve it with
\begin{CodeChunk}
\begin{CodeInput}
R> S <- Variable(n, n, PSD = TRUE)
R> obj <- log_det(S) - matrix_trace(S %*% Q)
R> constr <- list(sum(abs(S)) <= alpha)
R> prob <- Problem(Maximize(obj), constr)
R> result <- solve(prob, solver = "SCS")
\end{CodeInput}
\end{CodeChunk}
The \code{PSD = TRUE} argument to the \code{Variable} constructor restricts 
\code{S} to the positive semidefinite cone. In our objective, we use \cvxr{} 
functions for the log-determinant and trace. The expression \code{matrix\_trace(S
  \%*\% Q)} is equivalent to \code{sum(diag(S \%*\% Q))}, but the
former is preferred because it is more efficient than making nested
function calls. However, a standalone atom does not exist for the determinant,
so we cannot replace \code{log\_det(S)} with \code{log(det(S))}
since \code{det} is undefined for a `\code{Variable}' object.

\begin{figure}[t!]
  \centering
  \begin{subfigure}[b]{0.4\textwidth}
    \includegraphics[width=\textwidth]{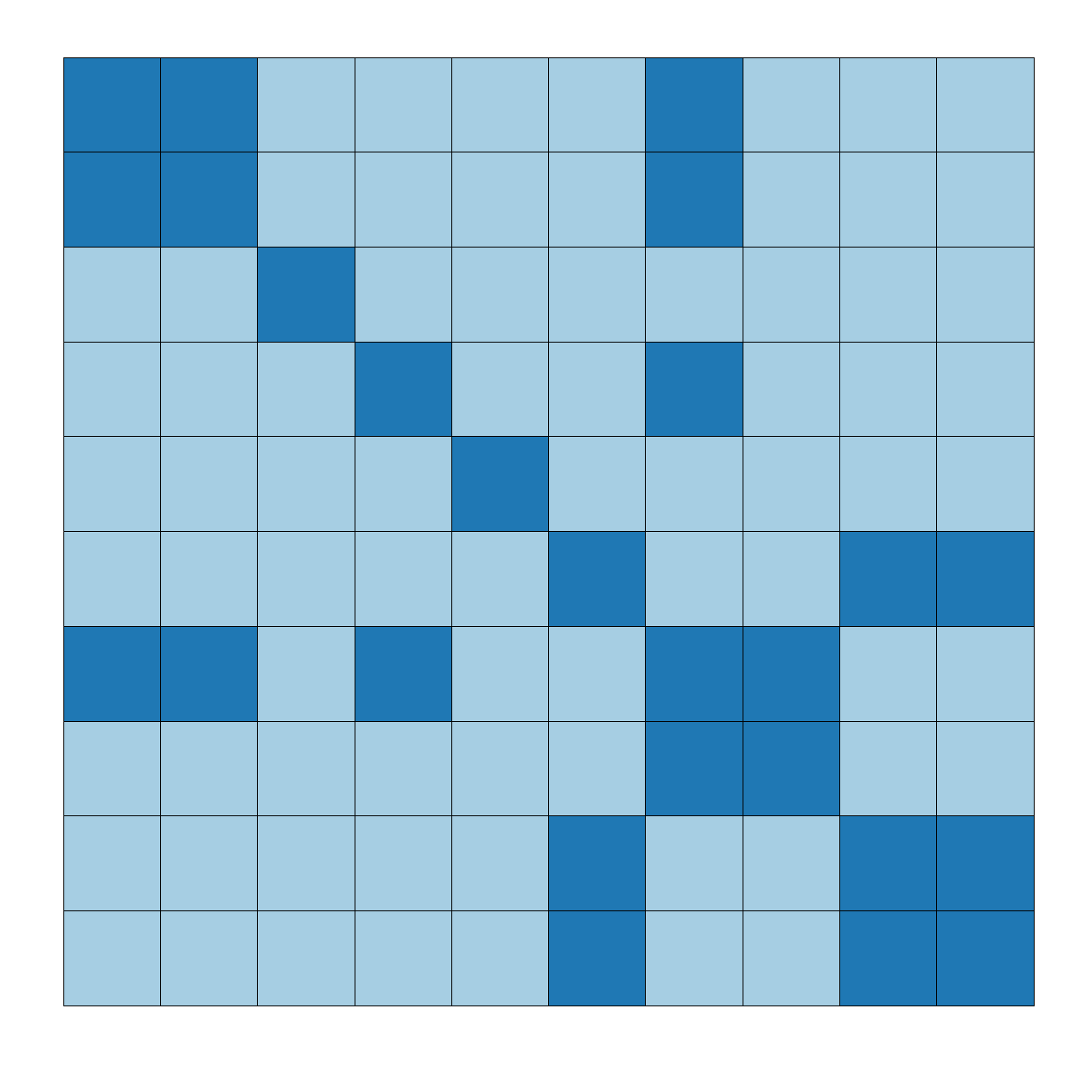}
    \caption{True inverse.}
  \end{subfigure}
  \begin{subfigure}[b]{0.4\textwidth}
    \includegraphics[width=\textwidth]{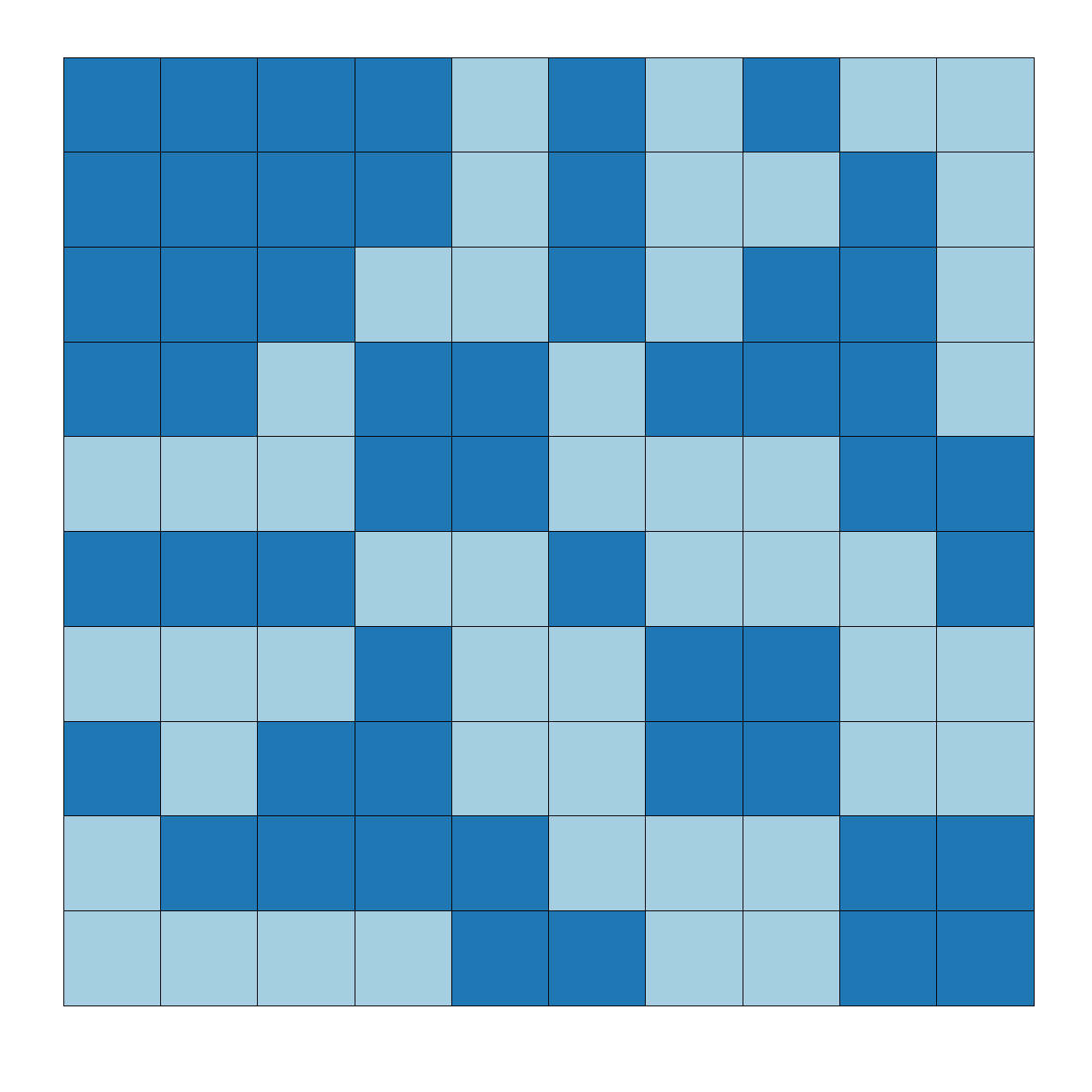}
    \caption{$\alpha = 10$.}
  \end{subfigure}
  \paragraph{\bigskip}
  \begin{subfigure}[b]{0.4\textwidth}
    \includegraphics[width=\textwidth]{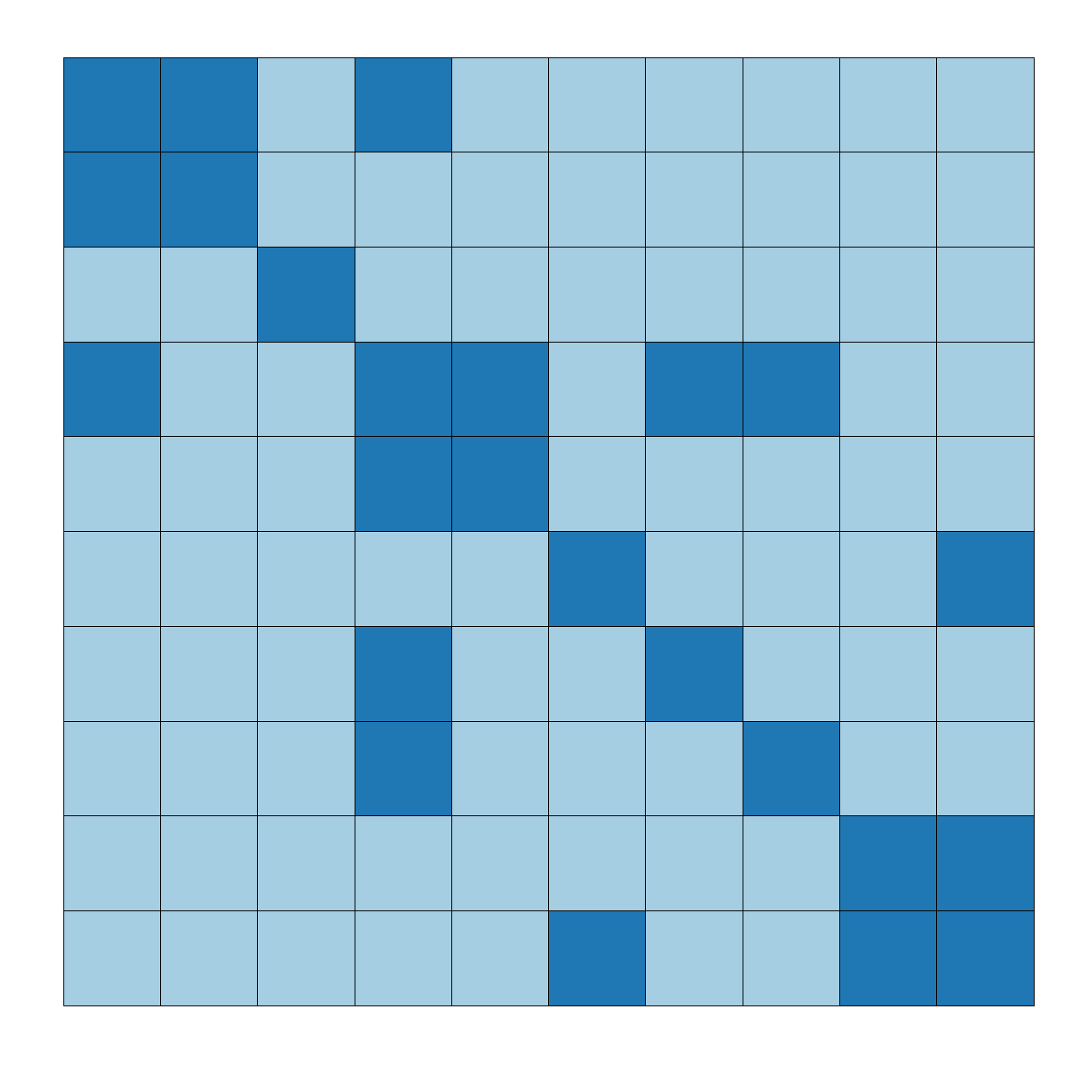}
    \caption{$\alpha = 4$.}
  \end{subfigure}
  \begin{subfigure}[b]{0.4\textwidth}
    \includegraphics[width=\textwidth]{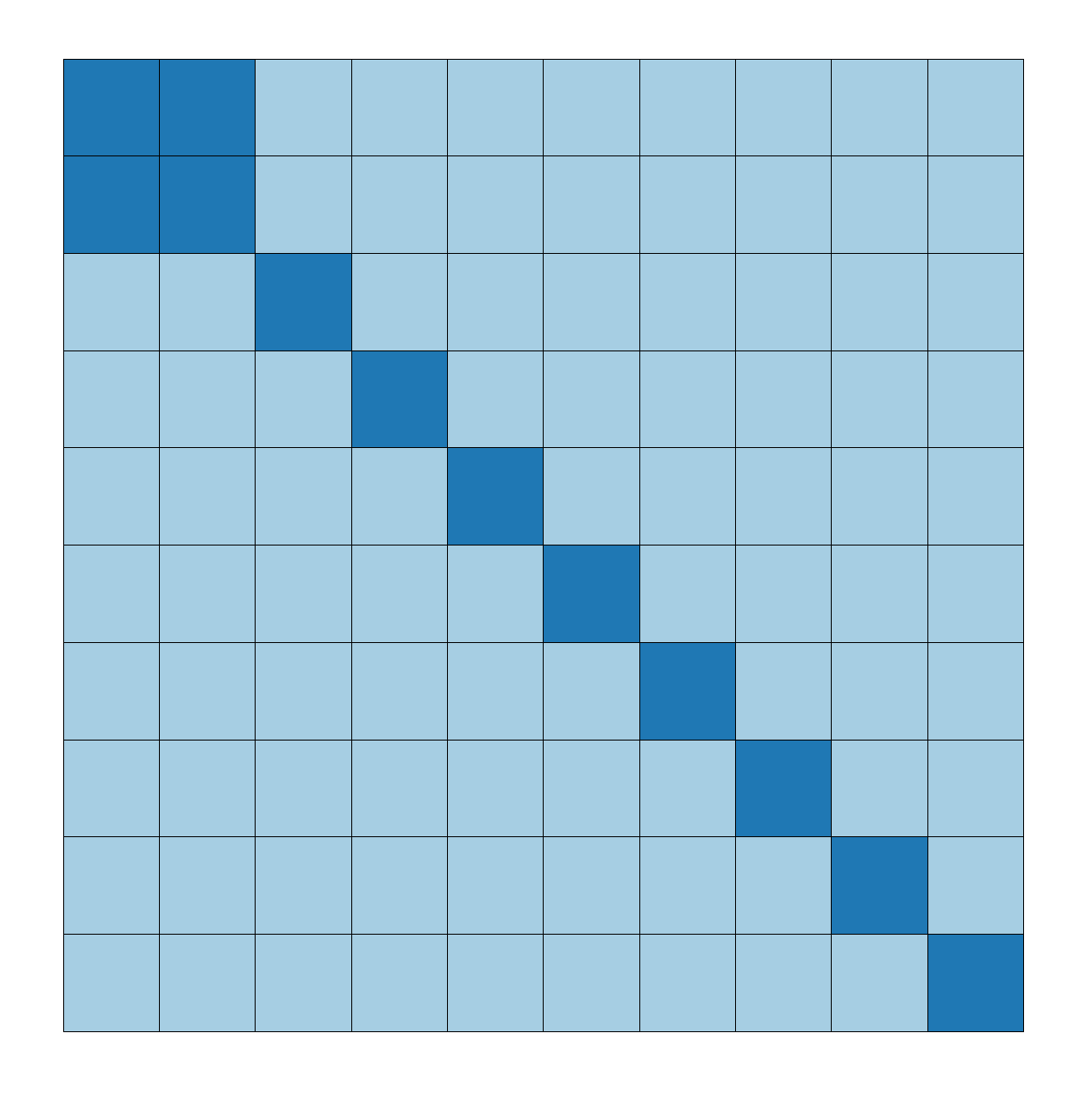}
    \caption{$\alpha = 1$.}
  \end{subfigure}
  \caption{Sparsity patterns for (a) inverse of true covariance
    matrix, and estimated inverse covariance matrices with (b)
    $\alpha = 10$, (c) $\alpha = 4$, and (d) $\alpha = 1$. The light
    blue regions indicate where $S_{ij} = 0$.}
  \label{fig:sparse-cov}
\end{figure}

Figure~\ref{fig:sparse-cov} depicts the solutions for a particular
dataset with $m = 1000, n = 10$, and $S$ containing 26\% non-zero
entries represented by the black squares in the top left image. The
sparsity of our inverse covariance estimate decreases for higher
$\alpha$, so that when $\alpha = 1$, most of the off-diagonal entries
are zero, while if $\alpha = 10$, over half the matrix is dense. At
$\alpha = 4$, we achieve the true percentage of non-zeros.

\subsubsection{Saturating hinges}
\label{ex:hinge}
The following example comes from work on saturating splines in
\citet{BoydHastie:2016}. Adaptive regression splines are commonly used
in statistical modeling, but the instability they exhibit beyond their
boundary knots makes extrapolation dangerous. One way to correct this
issue for linear splines is to require they {\em saturate}: remain
constant outside their boundary. This problem can be solved using a
heuristic that is an extension of lasso regression, producing a
weighted sum of hinge functions, which we call a {\em saturating
  hinge}.

For simplicity, consider the univariate case with $n = 1$. Assume we
are given knots $t_1 < t_2 < \cdots < t_k$ where each
$t_j \in \reals$. Let $h_j$ be a hinge function at knot $t_j$, i.e.,
$h_j(x) = \max(x-t_j,0)$, and define
$f(x) = w_0 + \sum_{j=1}^k w_jh_j(x)$. We want to solve
\[
  \begin{array}{ll} \underset{w_0,w}{\mbox{minimize}} & \sum_{i=1}^m \ell(y_i, f(x_i)) + \lambda\|w\|_1 \\
    \mbox{subject to} & \sum_{j=1}^k w_j = 0
\end{array}
\]
for variables $(w_0,w) \in \reals \times \reals^k$. The function
$\ell:\reals \times \reals \rightarrow \reals$ is the loss associated
with every observation, and $\lambda \geq 0$ is the penalty weight. In
choosing our knots, we set $t_1 = \min(x_i)$ and $t_k = \max(x_i)$ so
that by construction, the estimate $\hat f$ will be constant outside
$[t_1,t_k]$.

We demonstrate this technique on the bone density data for female
patients from \citet[Section 5.4]{ESL}. There are a total of $m = 259$
observations. Our response $y_i$ is the change in spinal bone density
between two visits, and our predictor $x_i$ is the patient's age. We
select $k = 10$ knots about evenly spaced across the range of $X$ and
fit a saturating hinge with squared error loss
$\ell(y_i, f(x_i)) = (y_i - f(x_i))^2$.

\begin{figure}[t!]
  \centering
  \begin{subfigure}[b]{0.49\textwidth}
    \includegraphics[width=\textwidth]{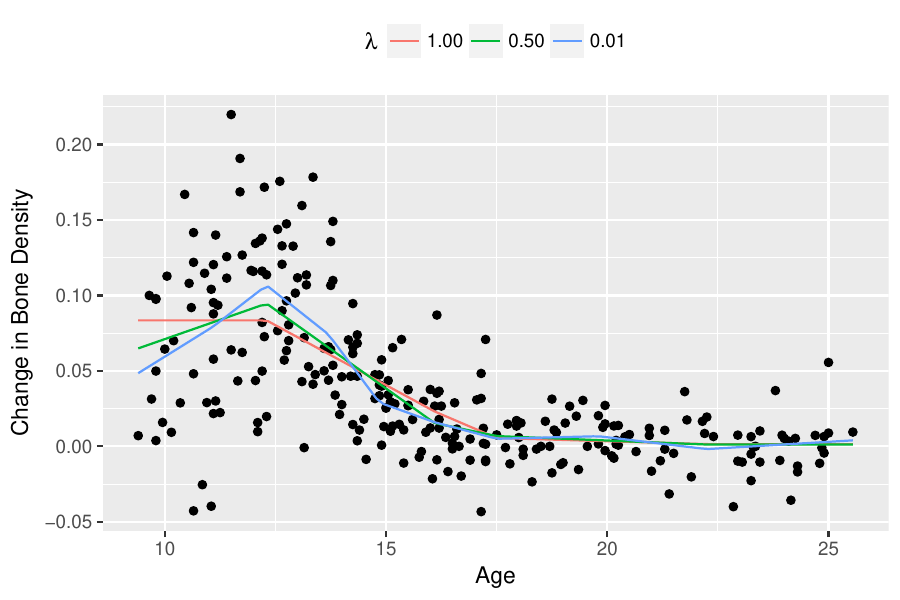}
    \caption{Squared error loss.}
    \label{fig:hinge-sse}
  \end{subfigure}
  \begin{subfigure}[b]{0.49\textwidth}
    \includegraphics[width=\textwidth]{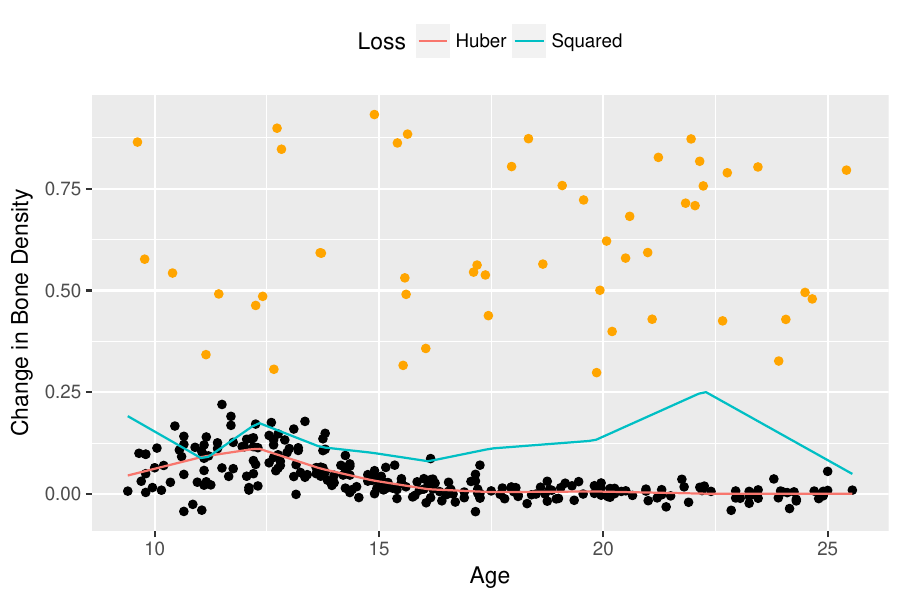}
    \caption{Saturating hinges with outliers.}
    \label{fig:hinge-out}
  \end{subfigure}
  \caption{(a) Saturating hinges fit to the change in bone density for
    female patients with $\lambda = 0.01$ (blue), $\lambda = 0.5$
    (green), and $\lambda = 1$ (red). (b) Hinges refit to the previous
    data with additional outliers (orange) using squared error (blue)
    and Huber loss (red).}
  \label{fig:hinge}
\end{figure}

In \proglang{R}, we first define the estimation and loss functions:
\begin{CodeChunk}
\begin{CodeInput}
R> f_est <- function(x, knots, w0, w) {
+    hinges <- sapply(knots, function(t) pmax(x - t, 0))
+    w0 + hinges %*% w
+  }
R> loss_obs <- function(y, f) (y - f)^2
\end{CodeInput}
\end{CodeChunk}
This allows us to easily test different losses and knot locations
later. The rest of the set-up is similar to previous examples. We
assume that \code{knots} is a \proglang{R} vector representing
$(t_1,\ldots,t_k)$.
\begin{CodeChunk}
\begin{CodeInput}
R> w0 <- Variable(1)
R> w <- Variable(k)
R> loss <- sum(loss_obs(y, f_est(X, knots, w0, w)))
R> reg <- lambda * p_norm(w, 1)
R> obj <- loss + reg
R> constr <- list(sum(w) == 0)
R> prob <- Problem(Minimize(obj), constr)
R> result <- solve(prob)
\end{CodeInput}
\end{CodeChunk}
The optimal weights are retrieved using separate calls, as shown
below.
\begin{CodeChunk}
\begin{CodeInput}
R> w0s <- result$getValue(w0)
R> ws <- result$getValue(w)
\end{CodeInput}
\end{CodeChunk}
We plot the fitted saturating hinges in Figure~\ref{fig:hinge-sse}. As
expected, when $\lambda$ increases, the spline exhibits less variation
and grows flatter outside its boundaries. The squared error loss works
well in this case, but as we saw previously in this section, the Huber
loss is preferred when the dataset contains large outliers. We can
change the loss function by simply redefining
\begin{CodeChunk}
\begin{CodeInput}
R> loss_obs <- function(y, f, M) huber(y - f, M)
\end{CodeInput}
\end{CodeChunk}
and passing an extra threshold parameter in when initializing
\code{loss}. In Figure~\ref{fig:hinge-out}, we have added 50 randomly
generated outliers to the bone density data and plotted the re-fitted
saturating hinges. For a Huber loss with $M = 0.01$, the resulting
spline is fairly smooth and follows the shape of the original data, as
opposed to the spline using squared error loss, which is biased
upwards by a significant amount.

\subsection{Nonparametric estimation}
\label{ex:nonpar}
\subsubsection{Log-concave distribution estimation}\label{ex:logconcave}
Let $n = 1$ and suppose $x_i$ are i.i.d.\ samples from a log-concave
discrete distribution on $\{0,\ldots,K\}$ for some
$K \in \integers_+$. Define $p_k := \Prob(X = k)$ to be the
probability mass function. One method for estimating
$(p_0,\ldots,p_K)$ is to maximize the log-likelihood function
subject to a log-concavity constraint \citep{mlelogcave}, i.e.,
\[
\begin{array}{ll}
\underset{p}{\mbox{maximize}} & \sum_{k=0}^K M_k\log p_k \\
\mbox{subject to} & p \geq 0, \quad \sum_{k=0}^K p_k = 1, \\
& p_k \geq \sqrt{p_{k-1}p_{k+1}}, \quad k = 1,\ldots,K-1,
\end{array}
\]
where $p \in \reals^{K+1}$ is our variable of interest and $M_k$
represents the number of observations equal to $k$, so that
$\sum_{k=0}^K M_k = m$. The problem as posed above is not
convex. However, we can transform it into a convex optimization
problem by defining new variables $u_k = \log p_k$ and relaxing the
equality constraint to $\sum_{k=0}^K p_k \leq 1$, since the latter
always holds tightly at an optimal solution. The result is
\[
\begin{array}{ll}
\underset{u}{\mbox{maximize}} & \sum_{k=0}^K M_k u_k \\
\mbox{subject to} & \sum_{k=0}^K e^{u_k} \leq 1, \\
& u_k - u_{k-1} \geq u_{k+1} - u_k, \quad k = 1,\ldots,K-1.
\end{array}
\]

\begin{figure}[t!]
  \centering
  \includegraphics[width=\textwidth]{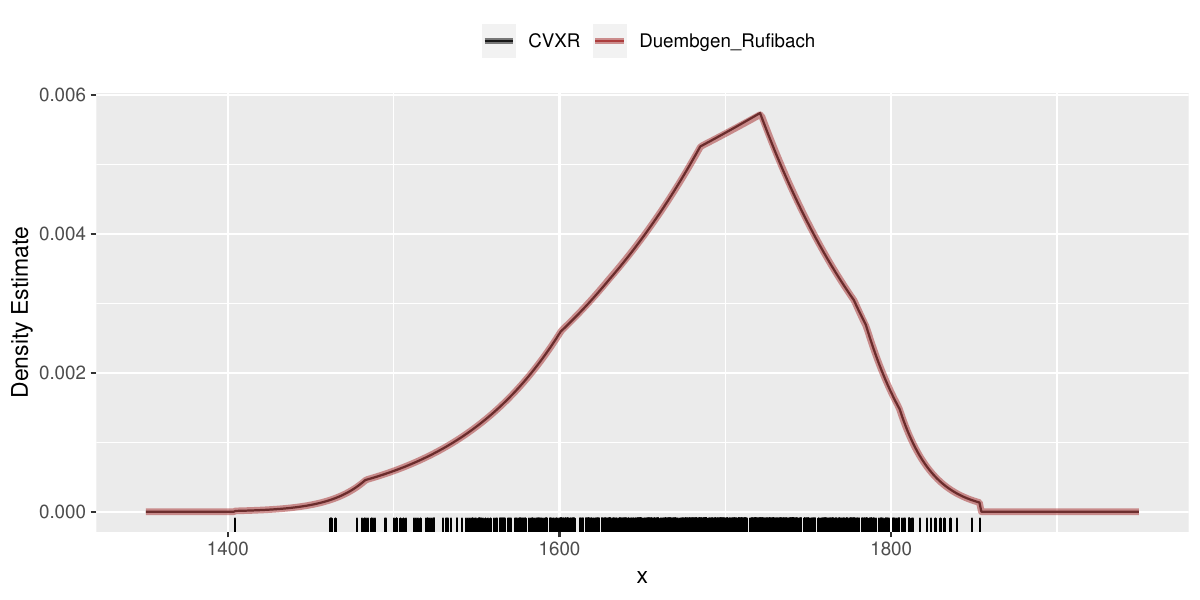}
  \caption{Log-concave estimation using the approach of
    \citet{deumbgen:rufibach:2011} and \cvxr{}.}
  \label{fig:log-concave}
\end{figure}

If \code{counts} is the \proglang{R} vector of $(M_0,\ldots,M_K)$, the code for our convex problem is
\begin{CodeChunk}
\begin{CodeInput}
R> u <- Variable(K+1)
R> obj <- t(counts) %*% u
R> constr <- list(sum(exp(u)) <= 1, diff(u[1:K)]) >= diff(u[2:(K+1)]))
R> prob <- solve(Maximize(obj), constr)
R> result <- solve(prob)
\end{CodeInput}
\end{CodeChunk}
Once the solver is finished, we can retrieve the probabilities
directly with
\begin{CodeChunk}
\begin{CodeInput}
R> pmf <- result$getValue(exp(u))
\end{CodeInput}
\end{CodeChunk}
The above line transforms the variables $u_k$ to
$e^{u_k}$ before calculating their resulting values. This is possible
because \code{exp} is a member of \cvxr{}'s library of atoms, so it
can operate directly on a `\code{Variable}' object such as \code{u}.

As an example, we consider the reliability data from
\mbox{\citet{deumbgen:rufibach:2011}} that was collected as part of a
consulting project at the Institute for Mathematical Statistics and
Actuarial Science, University of Bern \mbox{\citep{mlelogcave}}. The
dataset consists of $n =
786$ observations, and the goal is to fit a suitable distribution to
this sample that can be used for simulations. For various reasons
detailed in the paper, the authors chose a log-concave estimator,
which they implemented in the \proglang{R} package \pkg{logcondens}
\citep{deumbgen:rufibach:2011}. Figure~\ref{fig:log-concave} shows
that the curve obtained from the \cvxr{} code above matches their
results exactly.

\subsubsection{Survey calibration}\label{ex:calibration}

  Calibration is a widely used technique in survey sampling. Suppose
  $m$ sampling units in a survey have been assigned initial weights
  $d_i$ for $i = 1,\ldots,m$, and furthermore, there are $n$ auxiliary
  variables whose values in the sample are known. Calibration seeks to
  improve the initial weights $d_i$ by finding new weights $w_i$ that
  incorporate this auxiliary information while perturbing the initial
  weights as little as possible, i.e., the ratio $g_i = w_i/d_i$ must
  be close to one. Such reweighting improves precision of estimates
  \citep[Chapter 7]{lumley:2010}.

  Let $X \in \reals^{m \times n}$ be the matrix of survey samples, with each column corresponding to an auxiliary variable. Reweighting can be expressed as the optimization problem
\[
	\begin{array}{ll}
		\mbox{minimize} & \sum_{i=1}^m d_i\phi(g_i) \\
		\mbox{subject to} & A^\top g = r
	\end{array}
\]
with respect to $g \in \reals^m$, where
$\phi:\reals \rightarrow \reals$ is a strictly convex function with
$\phi(1) = 0$, $r \in \reals^n$ are the known
population totals of the auxiliary variables, and $A \in \reals^{m \times n}$ is related to $X$ by
$A_{ij} = d_iX_{ij}$ for $i = 1,\ldots,m$ and $j = 1,\ldots,n$. A common technique is raking, which uses the penalty function $\phi(g_i) = g_i\log(g_i) - g_i + 1$.

We illustrate with the California Academic Performance Index data in
the \pkg{survey} package \citep{Lumley:2004, lumley:2018}, which also
supplies facilities for calibration via the function
\code{calibrate}. Both the population dataset (\code{apipop}) and a
simple random sample of $m = 200$ (\code{apisrs}) are
provided. Suppose that we wish to reweight the observations in the
sample using known totals for two variables from the population:
\code{stype}, the school type (elementary, middle or high) and
\code{sch.wide}, whether the school met the yearly target or not. This
reweighting would make the sample more representative of the general
population.

\begin{table}[t!]
	\centering
\begin{tabular}{ccrrrr}
  \hline
  & & \multicolumn{2}{c}{\pkg{survey}} &\multicolumn{2}{c}{\cvxr{}}\\
  \cline{3-6}
  \multicolumn{1}{c}{School Type} & \multicolumn{1}{c}{Target Met?} & Weight & Frequency & Weight & Frequency \\ 
  \hline
E & Yes & 29.00 &  15 & 29.00 &  15 \\ 
H & No & 31.40 & 13 & 31.40 & 13 \\ 
M & Yes & 29.03 &  9 & 29.03 &  9 \\ 
E & No & 28.91 &  127 & 28.91 &  127 \\ 
H & Yes & 31.50 &   12 & 31.50 &   12 \\ 
M & No & 31.53 &  24 & 31.53 &  24 \\ 
  \hline
\end{tabular}
\caption{Raking weight estimates with \pkg{survey} package and \cvxr{}
  for California Academic Performance Index data.}
\label{tab:calib}
\end{table}

The code below solves the problem in \cvxr{}, where we have used a model matrix to generate the appropriate dummy variables for the two factor variables.
\begin{CodeChunk}
\begin{CodeInput}
R> m <- nrow(apisrs)
R> di <- apisrs$pw
R> formula <- ~ stype + sch.wide
R> r <- apply(model.matrix(object = formula, data = apipop), 2, sum)
R> X <- model.matrix(object = formula, data = apisrs)
R> A <- di * X
R> g <- Variable(m)
R> obj <- sum(di * (-entr(g) - g + 1))
R> constr <- list(t(A) %*% g == r)
R> prob <- Problem(Minimize(obj), constr)
R> result <- solve(prob)
R> w_cvxr <- di * result$getValue(g)
\end{CodeInput}
\end{CodeChunk}
Table~\ref{tab:calib} shows that the results are identical to those
obtained from \pkg{survey}. \cvxr{} can also accommodate other
penalty functions common in the survey literature, as well as additional
constraints.

\subsubsection{Nearly-isotonic and nearly-convex fits}
\label{ex:boot}

\begin{figure}[t!]
  \centering
  \begin{subfigure}{.5\textwidth}
    \centering
    \includegraphics[width=\linewidth]{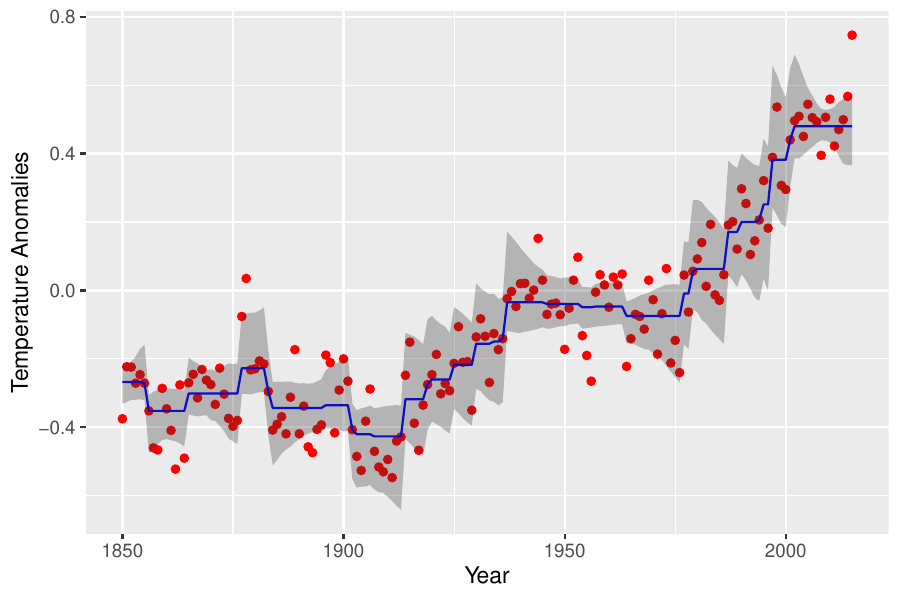}
    \caption{Nearly-isotonic.}
    \label{fig:near-iso}
  \end{subfigure}%
  \begin{subfigure}{.5\textwidth}
    \centering
    \includegraphics[width=\linewidth]{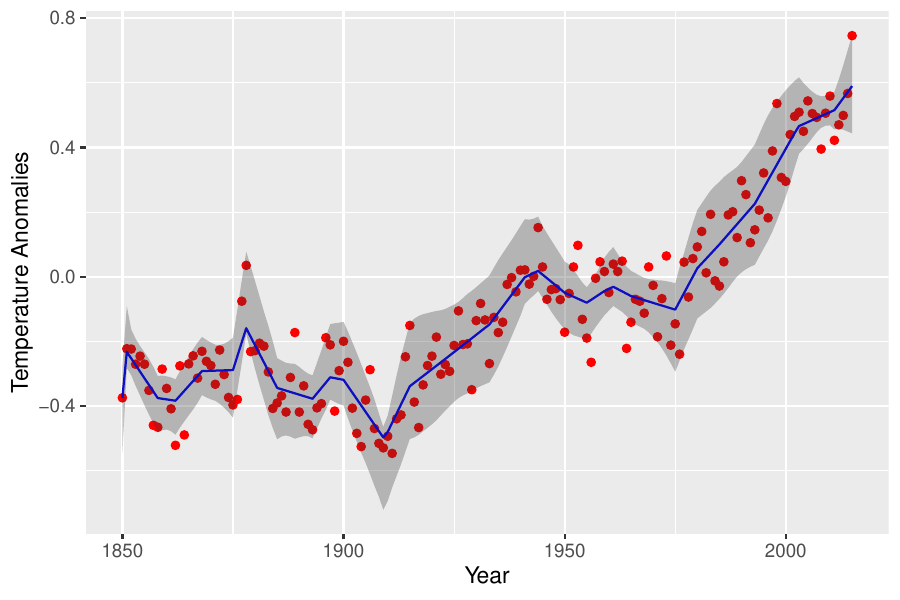}
    \caption{Nearly-convex.}
    \label{fig:near-convex}
  \end{subfigure}
  \caption{(a) A nearly-isotonic fit and (b) nearly-convex fit to
    global warming data on temperature anomalies for $\lambda = 0.44.$
    The 95\% normal confidence intervals are shown in gray using $R = 400$ and $R = 200$
    bootstrap samples, respectively.}
  \label{fig:near}
\end{figure}

Given a set of data points $y \in \reals^m$,
\citet{TibshiraniHoefling:2011} fit a nearly-isotonic approximation
$\beta \in \reals^m$ by solving
\[
\begin{array}{ll}
	\underset{\beta}{\mbox{minimize}} & \frac{1}{2}\sum_{i=1}^m (y_i - \beta_i)^2 + \lambda \sum_{i=1}^{m-1}(\beta_i - \beta_{i+1})_+,
\end{array}
\]
where $\lambda \geq 0$ is a penalty parameter and $x_+ =
\max(x,0)$. Our \cvxr{} formulation follows directly as shown below. The \code{pos} atom evaluates $x_+$ elementwise on the input expression.
\begin{CodeChunk}
\begin{CodeInput}
R> near_fit <- function(y, lambda) {
+    m <- length(y)
+    beta <- Variable(m)
+    penalty <- sum(pos(diff(beta)))
+    obj <- 0.5 * sum((y - beta)^2) + lambda * penalty
+    prob <- Problem(Minimize(obj))
+    result <- solve(prob)
+    result$getValue(beta)
+  }
\end{CodeInput}
\end{CodeChunk}
We demonstrate this technique on the global warming data provided by
the Carbon Dioxide Information Analysis Center (CDIAC). Our data
points are the annual temperature anomalies relative to the 1961--1990
mean. Combining \code{near_fit} with the \pkg{boot} package
\citep{boot}, we can obtain the standard errors and confidence
intervals for our estimate in just a few lines of code.
\begin{CodeChunk}
\begin{CodeInput}
R> near_fit_stat <- function(data, index, lambda) {
+    sample <- data[index, ]                  # Bootstrap sample of rows
+    sample <- sample[order(sample$year), ]   # Order ascending by year
+    near_fit(sample$annual, lambda)
+  }
R> boot.out <- boot(CDIAC, near_fit_stat, R = 400, lambda = 0.44)
\end{CodeInput}
\end{CodeChunk}
Figure~\ref{fig:near-iso} shows a nearly-isotonic fit with $\lambda = 0.44$ and 95\% normal confidence bands, which were generated using $R = 400$ bootstrap samples. The curve follows the data well, but exhibits choppiness in regions with a steep trend.

For a smoother curve, we can solve for the nearly-convex fit described in the same paper:
\[
\begin{array}{ll}
\underset{\beta}{\mbox{minimize}} & \frac{1}{2}\sum_{i=1}^m (y_i - \beta_i)^2 + \lambda \sum_{i=1}^{m-2}(\beta_i - 2\beta_{i+1} + \beta_{i+2})_+
\end{array}
\]
This replaces the first difference term with an approximation to the second derivative at $\beta_{i+1}$. In \cvxr{}, the only change necessary is the penalty line in \code{near_fit},
\begin{CodeChunk}
\begin{CodeInput}
R> penalty <- sum(pos(diff(beta, differences = 2)))
\end{CodeInput}
\end{CodeChunk}
The resulting curve is depicted in Figure~\ref{fig:near-convex} with 95\% confidence bands generated from $R = 200$ samples. Note the jagged staircase pattern has been smoothed out. We can easily extend this example to higher-order differences or lags by modifying the arguments to \code{diff}.

\subsection{Miscellaneous applications}

\subsubsection{Worst case covariance}\label{ex:worstcov}
Suppose we have i.i.d.\ samples $x_i \sim N(0,\Sigma)$ for
$i = 1,\ldots,m$ and want to determine the maximum covariance of
$y = w^\top x = \sum_{i=1}^m w_ix_i$, where $w \in \reals^m$ is a
given vector of weights. We are provided limited information on the
elements of $\Sigma$. For example, we may know the specific value or
sign of certain $\Sigma_{jk}$, which are represented by upper and
lower bound matrices $L$ and $U \in \reals^{n \times n}$, respectively
\citep[pp.\ 171--172]{BoydVandenberghe:2004}. This situation can arise
when calculating the worst-case risk of an investment portfolio
\citep{worstrisk}. Formally, our optimization problem is
\[
\begin{array}{ll} \underset{\Sigma}{\mbox{maximize}} & w^\top\Sigma w \\
\mbox{subject to} & \Sigma \in \symm_+^n, \quad L_{jk} \leq \Sigma_{jk} \leq U_{jk}, \quad j,k = 1,\ldots,n.
\end{array}
\]

Consider the specific case
\[
w = \left[\begin{array}{l} \phantom{-}0.1 \\ \phantom{-}0.2 \\ -0.05 \\ \phantom{-}0.1\end{array}\right], \quad
\Sigma = \left[\begin{array}{cccc}
0.2 & + & + & \pm \\
+ & 0.1 & - & - \\
+ & - & 0.3 & + \\
\pm & - & + & 0.1
\end{array}\right],
\]
where a $+$ means the element is non-negative, a $-$ means the element
is non-positive, and a $\pm$ means the element can be any real
number. In \cvxr{}, this semidefinite program is
\begin{CodeChunk}
\begin{CodeInput}
R> Sigma <- Variable(n, n, PSD = TRUE)
R> obj <- t(w) %*% Sigma %*% w
R> constr <- list(Sigma[1, 1] == 0.2, Sigma[1, 2] >= 0, Sigma[1, 3] >= 0,
+    Sigma[2, 2] == 0.1, Sigma[2, 3] <= 0, Sigma[2, 4] <= 0,
+    Sigma[3, 3] == 0.3, Sigma[3, 4] >= 0, Sigma[4, 4] == 0.1)
R> prob <- Problem(Maximize(obj), constr)
R> result <- solve(prob, solver = "SCS")
\end{CodeInput}
\end{CodeChunk}
Our result for this numerical case is
\[
	\Sigma = \left[\begin{array}{rrrr}
	0.2000 &  0.0967 &  0.0000 & 0.0762 \\
	0.0967 &  0.1000 & -0.1032 & 0.0000 \\
	0.0000 & -0.1032 &  0.3000 & 0.0041 \\
	0.0762 &  0.0000 &  0.0041 & 0.1000
	\end{array}\right]
\]

This example can be generalized to include arbitrary convex
constraints on $\Sigma$. Furthermore, if we have a target estimate for
the covariance, we can bound deviations from the target by
incorporating penalized slack variables into our optimization problem.

\begin{figure}[t!]
	\centering
		\includegraphics[width = 0.6\textwidth]{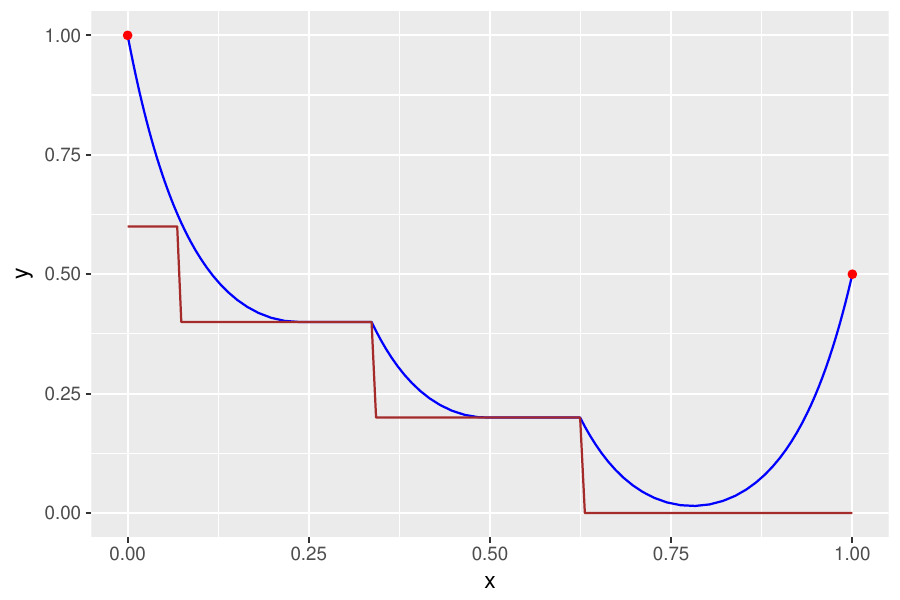}
	\caption{Solution of the catenary problem (blue) with a ground
		constraint (brown).}
	\label{fig:catenary}
\end{figure}

\subsubsection{Catenary problem}
\label{ex:catenary}
We consider a discretized version of the catenary problem in
\citet{catenary}. A chain with uniformly distributed mass hangs from
the endpoints $(0,1)$ and $(1,1)$ on a 2-D plane. Gravitational force
acts in the negative $y$ direction. Our goal is to find the shape of
the chain in equilibrium, which is equivalent to determining the
$(x,y)$ coordinates of every point along its curve when its potential
energy is minimized.

To formulate this as an optimization problem, we parameterize the
chain by its arclength and divide it into $m$ discrete links. The
length of each link must be no more than $h > 0$. Since mass is
uniform, the total potential energy is simply the sum of the
$y$-coordinates. Therefore, our problem is
\[
\begin{array}{ll} \underset{x,y}{\mbox{minimize}} & \sum_{i=1}^m y_i \\
\mbox{subject to} & x_1 = 0, \quad y_1 = 1, \quad x_m = 1, \quad y_m = 1 \\
& (x_{i+1} - x_i)^2 + (y_{i+1} - y_i)^2 \leq h^2, \quad i = 1,\ldots,m-1
\end{array}
\]
with variables $x \in \reals^m$ and $y \in \reals^m$. This basic
catenary problem has a well-known analytical solution
\citep{GelfandFomin:1963}, which we can easily verify with \cvxr{}.
\begin{CodeChunk}
\begin{CodeInput}
R> x <- Variable(m)
R> y <- Variable(m)
R> obj <- sum(y)
R> constr <- list(x[1] == 0, y[1] == 0, x[m] == 1, y[m] == 1,
+    diff(x)^2 + diff(y)^2 <= h^2)
R> prob <- Problem(Minimize(obj), constr)
R> result <- solve(prob)
\end{CodeInput}
\end{CodeChunk}
A more interesting situation arises when the ground is not flat. Let
$g \in \reals^m$ be the elevation vector (relative to the $x$-axis),
and suppose the right endpoint of our chain has been lowered by
$\Delta y_m = 0.5$. The analytical solution in this case would be
difficult to calculate. However, we need only add two lines to our
constraint definition,
\begin{CodeChunk}
\begin{CodeInput}
R> constr[[4]] <- (y[m] == 0.5)
R> constr <- c(constr, y >= g)
\end{CodeInput}
\end{CodeChunk}
to obtain the new result. Figure~\ref{fig:catenary} depicts the solution
of this modified catenary problem for $m = 101$ and $h = 0.02$. The
chain is shown hanging in blue, bounded below by the red staircase
structure, which represents the ground.

\subsubsection{Portfolio optimization}\label{ex:portfolio}
In this example, we solve the Markowitz portfolio problem under
various different constraints \citep{Markowitz:1952, Roy:1952,
  LoboFazelBoyd:2007}. We have $n$ assets or stocks in our portfolio
and must determine the amount of money to invest in each. Let $w_i$
denote the fraction of our budget invested in asset $i = 1,\ldots,m$,
and let $r_i$ be the returns (i.e., fractional change in price) over
the period of interest. We model returns as a random vector
$r \in \reals^n$ with known mean $\E[r] = \mu$ and covariance
$\VAR(r) = \Sigma$. Thus, given a portfolio $w \in \reals^n$, the
overall return is $R = r^\top w$.

\begin{figure}[t!]
  \centering
  \begin{subfigure}[b]{0.48\textwidth}
    \includegraphics[width=\textwidth]{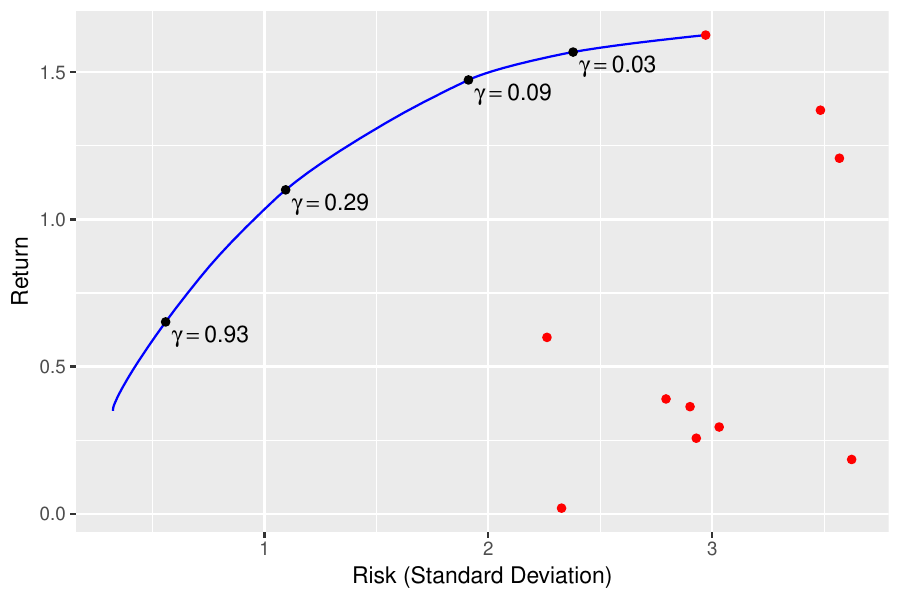}
    \caption{Risk-return curve.}
    \label{fig:portfolio-tradeoff}
  \end{subfigure}
  \begin{subfigure}[b]{0.48\textwidth}
    \includegraphics[width=\textwidth]{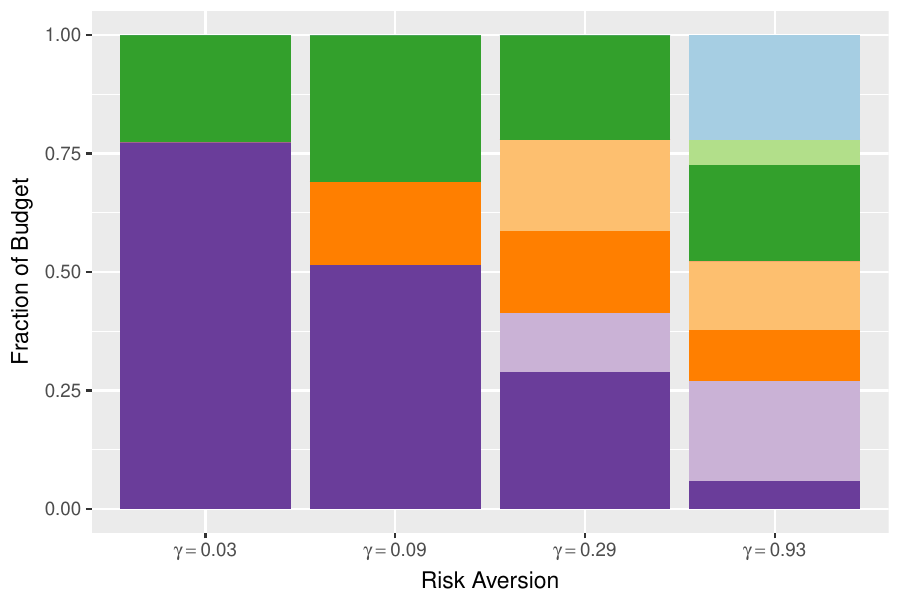}
    \caption{Asset portfolios.}
    \label{fig:portfolio-weights}
  \end{subfigure}
  \caption{(a) Risk-return trade-off curve for various
    $\gamma$. Portfolios that invest completely in one asset are
    plotted in red. (b) Fraction of budget invested in each
    asset.}
  \label{fig:portfolio}
\end{figure}

Portfolio optimization involves a trade-off between the expected
return $\E[R] = \mu^\top w$ and associated risk, which we take as the
return variance $\VAR(R) = w^\top\Sigma w$. Initially, we consider only
long portfolios, so our problem is
\[
	\begin{array}{ll} \underset{w}{\mbox{maximize}} & \mu^\top w - \gamma w^\top\Sigma w \\
	\mbox{subject to} & w \geq 0, \quad \sum_{i=1}^n w_i = 1,
	\end{array}
\]
where the objective is the risk-adjusted return and $\gamma > 0$ is a
risk aversion parameter.
\begin{CodeChunk}
\begin{CodeInput}
R> w <- Variable(n)
R> ret <- t(mu) %*% w
R> risk <- quad_form(w, Sigma)
R> obj <- ret - gamma * risk
R> constr <- list(w >= 0, sum(w) == 1)
R> prob <- Problem(Maximize(obj), constr)
R> result <- solve(prob)
\end{CodeInput}
\end{CodeChunk}
In this case, it is necessary to specify the quadratic form with \code{quad\_form} 
rather than the usual \code{t(w) \%*\% Sigma \%*\% w} because the latter will be 
interpreted by the \cvxr{} parser as a product of two affine terms and rejected 
for not being DCP.
We can obtain the risk and return by directly evaluating the value of
the separate expressions:
\begin{CodeChunk}
\begin{CodeInput}
R> result$getValue(risk)
R> result$getValue(ret)
\end{CodeInput}
\end{CodeChunk}
Figure~\ref{fig:portfolio-tradeoff} depicts the risk-return trade-off
curve for $n = 10$ assets and $\mu$ and $\Sigma^{1/2}$ drawn from a
standard normal distribution. The $x$-axis represents the standard
deviation of the return. Red points indicate the result from investing
the entire budget in a single asset. As $\gamma$ increases, our
portfolio becomes more diverse (Figure~\ref{fig:portfolio-weights}),
reducing risk but also yielding a lower return.

Many variations on the classical portfolio problem exist. For
instance, we could allow long and short positions, but impose a
leverage limit $\|w\|_1 \leq L^{\max}$ by changing
\begin{CodeChunk}
\begin{CodeInput}
R> constr <- list(p_norm(w, 1) <= Lmax, sum(w) == 1)
\end{CodeInput}
\end{CodeChunk}
An alternative is to set a lower bound on the return and minimize just
the risk. To account for transaction costs, we could add a term to the
objective that penalizes deviations of $w$ from the previous
portfolio. These extensions and more are described in
\citet{BoydBusseti:2017}. The key takeaway is that all of these convex
problems can be easily solved in \cvxr{} with just a few alterations
to the code above.

\subsubsection{Kelly gambling}\label{ex:kelly}
In Kelly gambling \citep{kelly}, we are given the opportunity to bet on
$n$ possible outcomes, which yield a random non-negative return of
$r \in \reals_+^n$. The return $r$ takes on exactly $K$ values
$r_1,\ldots,r_K$ with known probabilities $\pi_1,\ldots,\pi_K$. This
gamble is repeated over $T$ periods. In a given period $t$, let
$b_i \geq 0$ denote the fraction of our wealth bet on outcome
$i$. Assuming the $n$th outcome is equivalent to not wagering (it
returns one with certainty), the fractions must satisfy
$\sum_{i=1}^n b_i = 1$. Thus, at the end of the period, our cumulative
wealth is $w_t = (r^\top b)w_{t-1}$. Our goal is to maximize the average
growth rate with respect to $b \in \reals^n$:
\[
\begin{array}{ll} \underset{b}{\mbox{maximize}} & \sum_{j=1}^K \pi_j\log(r_j^\top b) \\
	\mbox{subject to} & b \geq 0, \quad \sum_{i=1}^n b_i = 1.
\end{array}
\]
In the following code, \code{rets} is the $K$ by $n$ matrix of
possible returns with rows $r_j$, while \code{ps} is the vector of
return probabilities $(\pi_1,\ldots,\pi_K)$.
\begin{CodeChunk}
\begin{CodeInput}
R> b <- Variable(n)
R> obj <- t(ps) %*% log(rets %*% b)
R> constr <- list(b >= 0, sum(b) == 1)
R> prob <- Problem(Maximize(obj), constr)
R> result <- solve(prob)
\end{CodeInput}
\end{CodeChunk}
We solve the Kelly gambling problem for $K = 100$ and $n = 20$. The
probabilities $\pi_j \sim \mbox{Uniform}(0,1)$, and the potential
returns $r_{ji} \sim \mbox{Uniform}(0.5,1.5)$ except for $r_{jn} = 1$, 
which represents the payoff from not wagering. With an initial wealth 
of $w_0 = 1$, we simulate the growth trajectory of our Kelly optimal 
bets over $P = 100$ periods, assuming returns are i.i.d.\ over time.
\begin{CodeChunk}
\begin{CodeInput}
R> bets <- result$getValue(b)
R> idx <- sample.int(K, size = P, probs = ps, replace = TRUE)
R> winnings <- rets[idx,] %*% bets
R> wealth <- w0 * cumprod(winnings)
\end{CodeInput}
\end{CodeChunk}
For comparison, we also calculate the trajectory for a na\"{i}ve
betting scheme, which holds onto 15\% of the wealth at the beginning
of each period and divides the other 85\% over the bets in direct
proportion to their expect returns.

\begin{figure}[t!]
\centering
\includegraphics[width=0.7\textwidth]{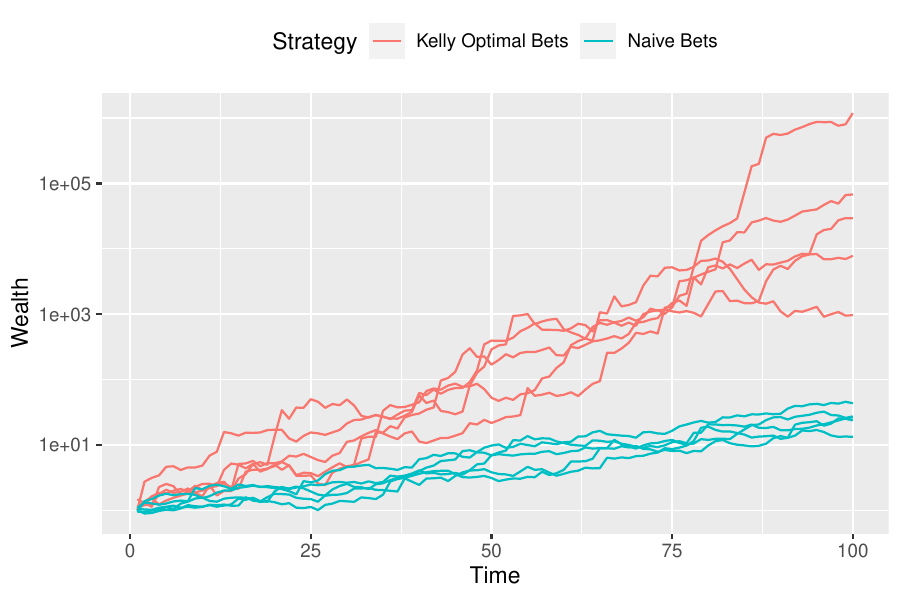}
\caption{Wealth trajectories for the Kelly optimal bets (red) and
	na\"{i}ve bets (cyan). The na\"{i}ve betting scheme holds onto 15\%
	of the wealth and splits the rest in direct proportion to the
	expected returns.}
\label{fig:kelly-growth}
\end{figure}

Growth curves for five independent trials are plotted in Figure
\ref{fig:kelly-growth}. Red lines represent the wealth each period from
the Kelly bets, while cyan lines are the result of the na\"{i}ve
bets. Clearly, Kelly optimal bets perform better, producing greater
net wealth by the final period. However, as observed in some
trajectories, wealth tends to drop by a significant amount before
increasing eventually. One way to reduce this drawdown risk is to add
a convex constraint as proposed in \citet[Section 5.3]{RCK},
\[
	\log\left(\sum_{j=1}^K \exp(\log\pi_j - \lambda \log(r_j^\top b))\right) \leq 0,
\]
where $\lambda \geq
0$ is the risk-aversion parameter. With \cvxr{}, this can be
accomplished in a single line using the \code{log\_sum\_exp}
atom. Other extensions like wealth goals, betting restrictions, and
VaR/CVaR bounds are also readily incorporated.
\newpage
\subsubsection{Channel capacity}\label{ex:channel}
The following problem comes from an exercise in \citet[pp.\
207--208]{BoydVandenberghe:2004}. Consider a discrete memoryless
communication channel with input $X(t) \in \{1,\ldots,n\}$ and output
$Y(t) \in \{1,\ldots,m\}$ for $t = 1,2,\ldots$. The relation between
the input and output is given by a transition matrix
$P \in \reals_+^{m \times n}$ with
\[
P_{ij} = \Prob(Y(t) = i|X(t) = j), \quad i = 1,\ldots,m, \quad j = 1,\ldots,n.
\]
Assume that $X$ has a probability distribution denoted by
$x \in \reals^n$, i.e., $x_j = \Prob(X(t) = j)$ for $j = 1,\ldots,n$. A
famous result by \citet{ShannonWeaver:1949} states that the
channel capacity is found by maximizing the mutual information between
$X$ and $Y$,
\[
I(X,Y) = \sum_{j=1}^n x_j \sum_{i=1}^m P_{ij}\log_2P_{ij} - \sum_{i=1}^m y_i\log_2y_i,
\]
where $y = Px$ is the probability distribution of $Y$. Since $I$ is
concave, this is equivalent to solving the convex optimization problem
\[
\begin{array}{ll} \underset{x,y}{\mbox{maximize}} & \sum_{j=1}^n x_j \sum_{i=1}^m P_{ij}\log P_{ij} - \sum_{i=1}^m y_i\log y_i \\
\mbox{subject to} & x \geq 0, \quad \sum_{i=1}^m x_i = 1, \quad y = Px
\end{array}
\]
for $x \in \reals^n$ and $y \in \reals^m$. The associated code in
\cvxr{} is
\begin{CodeChunk}
\begin{CodeInput}
R> x <- Variable(n)
R> y <- P %*% x
R> c <- apply(P * log2(P), 2, sum)
R> obj <- t(c) %*% x + sum(entr(y))
R> constr <- list(sum(x) == 1, x >= 0)
R> prob <- Problem(Maximize(obj), constr)
R> result <- solve(prob)
\end{CodeInput}
\end{CodeChunk}
The channel capacity is simply the optimal objective,
\code{result\$value}.

\subsubsection{Fastest mixing Markov chain}\label{ex:fmmc}
This example is derived from the results in \citet[Section 2]{FMMC}. Let
$\mathcal{G} = (\mathcal{V}, \mathcal{E})$ be a connected graph with
vertices $\mathcal{V} = \{1,\ldots,n\}$ and edges
$\mathcal{E} \subseteq \mathcal{V} \times \mathcal{V}$. Assume that
$(i,i) \in \mathcal{E}$ for all $i = 1,\ldots,n$, and
$(i,j) \in \mathcal{E}$ implies $(j,i) \in \mathcal{E}$. Under these
conditions, a discrete-time Markov chain on $\mathcal{V}$ will have
the uniform distribution as one of its equilibrium distributions. We
are interested in finding the Markov chain, i.e., constructing the
transition probability matrix $P \in \reals_+^{n \times n}$, that
minimizes its asymptotic convergence rate to the uniform
distribution. This is an important problem in Markov chain Monte Carlo
(MCMC) simulations, as it directly affects the sampling efficiency of
an algorithm.

The asymptotic rate of convergence is determined by the second largest
eigenvalue of $P$, which in our case is
$\mu(P) := \sigma_{\max}(P - \frac{1}{n}\ones\ones^\top)$ where
$\sigma_{\max}(A)$ denotes the maximum singular value of $A$. As $\mu(P)$ decreases, the mixing rate increases and the Markov chain converges faster to equilibrium. Thus, our optimization problem is
\[
\begin{array}{ll}
\underset{P}{\mbox{minimize}} & \sigma_{\max}(P - \frac{1}{n}\ones\ones^\top) \\
\mbox{subject to} & P \geq 0, \quad P\ones = \ones, \quad P = P^\top \\
& P_{ij} = 0, \quad (i,j) \notin \mathcal{E}.
\end{array}
\]
The element $P_{ij}$ of our transition matrix is the probability of
moving from state $i$ to state $j$. Our assumptions imply that $P$ is
non-negative, symmetric, and doubly stochastic. The last constraint
ensures transitions do not occur between unconnected vertices.

\begin{figure}[t!]
\centering
\begin{subfigure}[b]{0.48\textwidth}
\includegraphics[width=\textwidth]{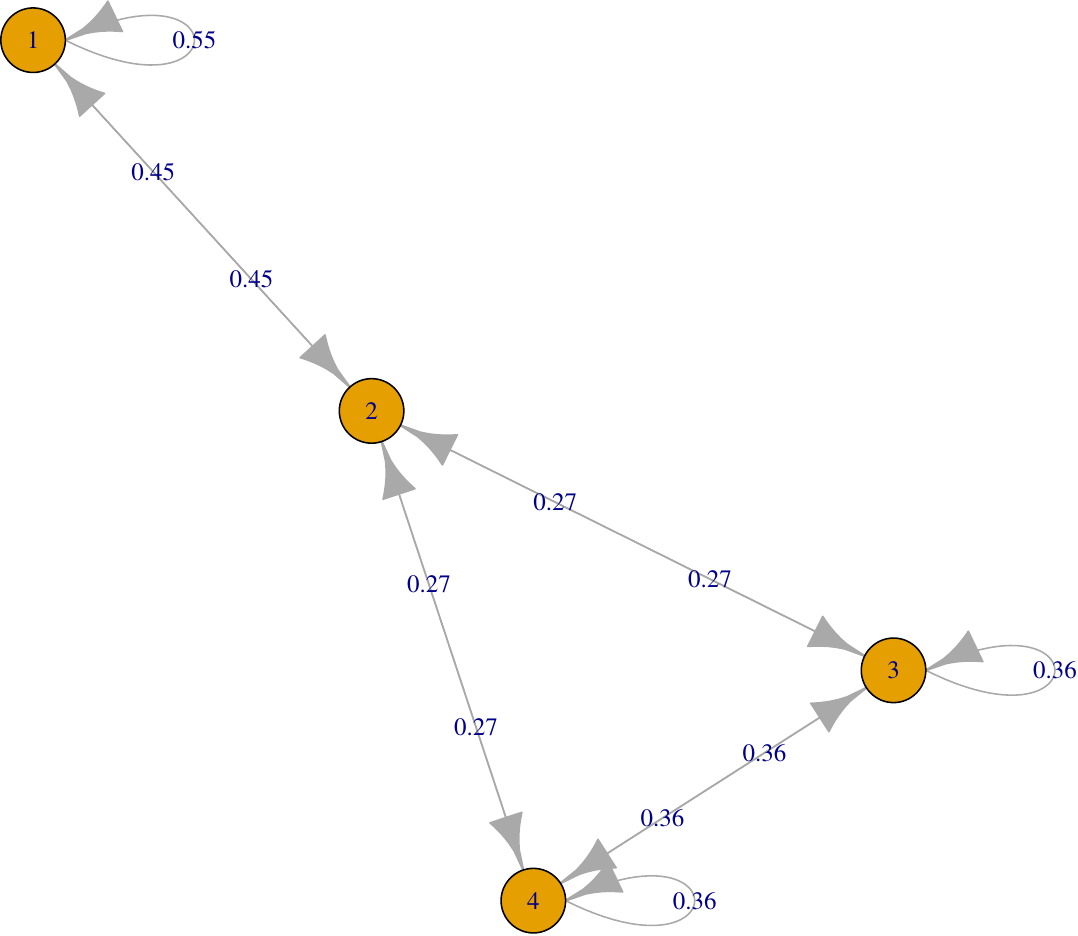}
\caption{Triangle + 1 edge.}
\label{fig:fmmc-triangle}
\end{subfigure}
\begin{subfigure}[b]{0.48\textwidth}
\includegraphics[width=\textwidth]{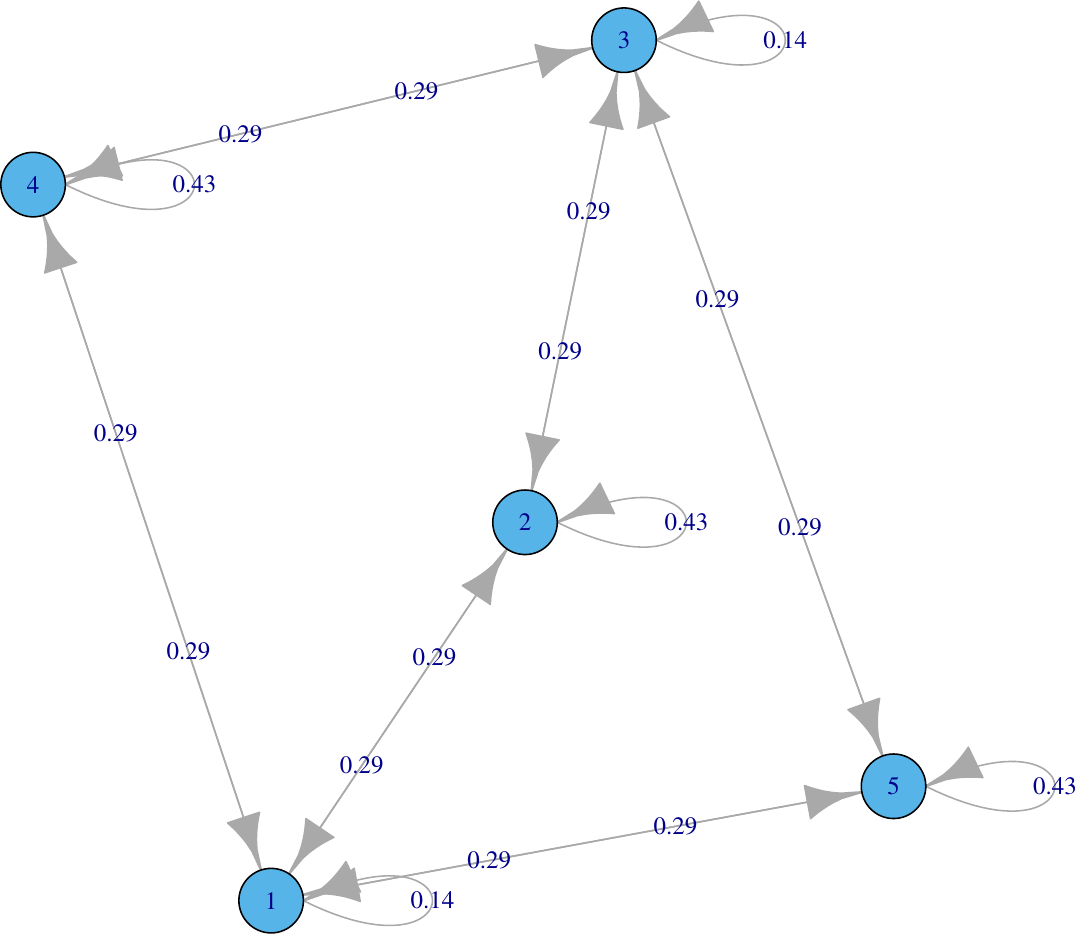}
\caption{Bipartite 2 + 3.}
\label{fig:fmmc-bipartite}
\end{subfigure}
\caption{Markov chains with transition probabilities that achieve the fastest mixing rate.}
\label{fig:fmmc}
\end{figure}

The function $\sigma_{\max}$ is convex, so this problem is solvable
in \cvxr{}. For instance, the code for the Markov chain in Figure
\ref{fig:fmmc-triangle} is
\begin{CodeChunk}
\begin{CodeInput}
R> P <- Variable(n, n)
R> ones <- matrix(1, nrow = n, ncol = 1)
R> obj <- sigma_max(P - 1/n)
R> constr1 <- list(P >= 0, P %*% ones == ones, P == t(P))
R> constr2 <- list(P[1, 3] == 0, P[1, 4] == 0)
R> prob <- Problem(Minimize(obj), c(constr1, constr2))
R> result <- solve(prob, solver = "SCS")
\end{CodeInput}
\end{CodeChunk}
where we have set $n = 4$. We could also have specified
$P\ones = \ones$ with \code{sum\_entries(P, 1) == 1}, which uses the
\code{sum\_entries} atom to represent the row sums.

It is easy to extend this example to other Markov chains. To change
the number of vertices, we would simply modify \code{n}, and to add
or remove edges, we need only alter the constraints in
\code{constr2}. For instance, the bipartite chain in Figure
\ref{fig:fmmc-bipartite} is produced by setting $n = 5$ and
\begin{CodeChunk}
\begin{CodeInput}
R> constr2 <- list(P[1, 3] == 0, P[2, 4] == 0, P[2, 5] == 0, P[4, 5] == 0)
\end{CodeInput}
\end{CodeChunk}
\section{Implementation}\label{implement}

\cvxr{} represents the atoms, variables, constraints, and other parts
of an optimization problem using \proglang{S}4 class
objects. \proglang{S}4 enables us to overload standard mathematical
operations so \cvxr{} combines seamlessly with native \proglang{R}
code and other packages. When an operation is invoked on a variable, a
new object is created that represents the corresponding expression
tree with the operator as the root node and the arguments as
leaves. This tree grows automatically as more elements are added,
allowing us to encapsulate the structure of an objective function or
constraint.

Once the user calls \code{solve}, DCP verification occurs. \cvxr{}
traverses the expression tree recursively, determining the sign and
curvature of each sub-expression based on the properties of its
component atoms. If the problem is deemed compliant, it is transformed
into an equivalent cone program using graph implementations of convex
functions \citep{GrantBoydYe:2006}. Then, \cvxr{} passes the problem's
description to the \pkg{CVXcanon} \proglang{C++} library \citep{cvxcanon},
which generates data for the cone program, and sends this data to the
solver-specific \proglang{R} interface. The solver's results are
returned to the user in a list. This object-oriented design and
infrastructure were largely borrowed from \pkg{CVXPY}. 

\cvxr{} interfaces with the open-source cone solvers \pkg{ECOS}
\citep{ECOS} and \pkg{SCS} \citep{SCS} through their respective
\proglang{R} packages. \pkg{ECOS} is an interior-point solver, which
achieves high accuracy for small and medium-sized problems, while
\pkg{SCS} is a first-order solver that is capable of handling larger
problems and semidefinite constraints. As noted by \citet[Section
I.A]{ECOS}, first-order methods can be slow if the problem is not well
conditioned or if it has a feasible set that does not allow for an
efficient projection, while interior-point methods have a convergence
rate that is independent of the problem data and the particular
feasible set. {Furthermore, starting from version 0.99, \cvxr{} also 
provides support for the commercial solvers \pkg{MOSEK} \mbox{\citep{MOSEK}} 
and \pkg{GUROBI} \mbox{\citep{GUROBI}} through binary \proglang{R} packages 
published by the respective vendors.} 
It is not difficult to connect
additional solvers so long as the solver has an API that can
communicate with \proglang{R}. Users who wish to employ a custom
solver may obtain the canonicalized data {for a problem and solver
combination} directly with {\code{get\_problem\_data(problem,
solver)}}. {When more than one solver is capable of solving a
problem, the \code{solver} argument to the \code{solve} function can
be used to indicate a preference. Available solvers, depending on
installed packages in a session, are returned via
\code{installed\_solvers()}. Interested users should consult
tutorial examples on the web page \url{https://cvxr.rbind.io} for
further guidance.}

We have provided a large library of atoms, which should be sufficient
to model most convex optimization problems. However, it is possible
for a sophisticated user to incorporate new atoms into this
library. The process entails creating a \proglang{S}4 class for the
atom, overloading methods that characterize its DCP properties, and
representing its graph implementation as a list of linear operators
that specify the corresponding feasibility problem. {For instance, the
  absolute value function $f(x) = |x|$ is represented by the
  `\code{Abs}' class, which inherits from `\code{Atom}'. We defined its
  curvature by overloading the \proglang{S}4 method
  \code{is\_atom\_convex}, used in the DCP verification step, to
  return \code{TRUE} when called on an `\code{Abs}' object. Then, we
  derived the graph form of the absolute value to be
  $f(x) = \inf\{ t| -t \leq x \leq t \}$. This form's objective and
  constraints were coded into lists in the atom's
  \code{graph\_implementation} function.} A full mathematical
exposition may be found in \citet[Section 10]{GrantBoydYe:2006}. In
general, we suggest users try to reformulate their optimization
problem first before attempting to add a novel atom.

\subsection{Speed considerations}\label{speed}

{Usually}, \cvxr{} will be slower than {a direct call to a solver},
because in the latter case, the user would have {already }done the job
of translating a mathematical problem into code and constraints
{ingestible by} the solver. \cvxr{} does this translation for the user
starting from a DCP formulation of the problem by walking the abstract
syntax tree, which represents the canonicalized objectives and
constraints, and building appropriate matrix structures for the
solver. The matrix data are passed to a compatible solver using either
\pkg{Rcpp} \citep{rcpp} or calls to a solver-specific \proglang{R}
package. \cvxr{} stores data in sparse matrices, thereby allowing
large problems to be specified. However, the restrictions imposed by
\proglang{R} on sparse matrices \citep{bates:martin:2018} still apply:
each dimension cannot exceed the integer limit of $2^{31}-1$.

Currently, the canonicalization and construction of data in
\proglang{R} for the solver dominates computation time, particularly
for complex expressions that involve indexing into individual elements 
of a matrix or vector. Using available \cvxr{} functions for
vectorized operations provides substantial speed improvements.

\cvxr{} also provides a \texttt{Parameter} object that can be combined
with warm starts, if such an option is available in the solver. A 
\texttt{Parameter} is a constant expression whose value can be modified 
after a \texttt{Problem} is created. This can yield significant reductions 
in computation time when solving a family of parametrized problems. The 
code below exploits warm starts to solve a lasso problem with two 
different values of the penalization parameter $\lambda$.

\begin{CodeChunk}
\begin{CodeInput}
R> beta <- Variable(n)
R> lambda <- Parameter(pos = TRUE)
R> obj <- 0.5 * sum((y - X %*% beta)^2) + lambda * p_norm(beta, 1)
R> constr <- list(beta >= 0)
R> prob <- Problem(Minimize(obj), constr)
R> value(lambda) <- 1   # First value of lambda
R> result <- solve(prob, solver = "OSQP")
R> value(lambda) <- 2   # Second value of lambda, warm start
R> result <- solve(prob, solver = "OSQP", warm_start = TRUE)
\end{CodeInput}
\end{CodeChunk}

On a commodity Macintosh laptop, with $X \in \reals^{2000 \times 500}$ and 
$y \in \reals^{2000}$, the first solution took 7.153 seconds, while the second 
took only 0.763 seconds.

\section{Conclusion}\label{conclusion}

Convex optimization plays an essential role in many fields,
particularly machine learning and statistics. \cvxr{} provides an
object-oriented language with which users can easily formulate,
modify, and solve a broad range of convex optimization problems. While
other \proglang{R} packages may perform faster on a subset of these
problems, \cvxr{}'s advantage is its flexibility and simple intuitive
syntax, making it an ideal tool for prototyping new models for which
custom \proglang{R} code does not exist. For more information, see the
official web page of the package on the Comprehensive \proglang{R}
Archive Network (CRAN) at
\url{https://CRAN.R-project.org/package=CVXR} and documentation.

\section*{Acknowledgments}
The authors would like to thank Trevor Hastie, Robert Tibshirani, John
Chambers, and David Donoho for their thoughtful advice and comments on
this project. {The authors also thank the referees
  for suggesting improvements and drawing our attention to some
  references.} We are grateful to Steven Diamond, John Miller, and
Paul Kunsberg Rosenfield for their contributions to the software's
development. In particular, we are indebted to Steven for his work on
\pkg{CVXPY}. Most of \cvxr{}'s code, documentation, and examples were
ported from his \proglang{Python} library.

Anqi Fu's research was
supported by the Stanford Graduate Fellowship and DARPA X-DATA
program. Balasubramanian Narasimhan's work was supported by the
Clinical and Translational Science Award 1UL1 RR025744 for the
Stanford Center for Clinical and Translational Education and Research
(Spectrum) from the National Center for Research Resources, National
Institutes of Health.

\bibliography{cvxr_refs}
\newpage
\appendix

\begin{sidewaystable}
  \small
  \centering
  \begin{tabular}{p{1.85in}p{1.6in}p{.9in}p{.8in}p{.6in}p{1in}}      
    \hline
    Function & Meaning & Domain & Sign & Curvature & Monotonicity\\
    \hline
     \code{geo\_mean(x)} \newline \code{geo\_mean(x, p)} \newline \(p \in \reals_+^n\), \(p \neq 0\) & \(x_1^{1/n} \cdots x_n^{1/n}\) \(\left(x_1^{p_1} \cdots x_n^{p_n}\right)^{\frac{1}{\mathbf{1}^\top p}}\) & \(x \in \reals_+^n\) & $+$ & concave & $\nearrow$\\
      \hline
      \code{harmonic\_mean(x)} & \(\frac{n}{\frac{1}{x_1} + \cdots + \frac{1}{x_n}}\) & \(x \in \mathbf{R}^n_{+}\) & $+$ & concave & $\nearrow$\\
      \hline
      \code{lambda\_max(X)} & \(\lambda_{\max}(X)\) & \(X \in
                                                      \mathbf{S}^n\) &
                                                                       $\pm$  & convex & none\\
      \hline
      \code{lambda\_min(X)} & \(\lambda_{\min}(X)\) & \(X \in \mathbf{S}^n\) &  & concave & none\\
      \hline
      \code{lambda\_sum\_largest(X, k)} \newline \(k = 1,\ldots, n\) & sum of $k$
                                                       largest eigenvalues of $X$ & \(X \in\mathbf{S}^{n}\) & $\pm$ & convex & none\\
      \hline
      \code{lambda\_sum\_smallest(X, k)} \newline \(k = 1,\ldots, n\) & sum of $k$
                                                        smallest eigenvalues of $X$ & \(X \in\mathbf{S}^{n}\) & $\pm$ & concave & none\\
      \hline
      \code{log\_det(X)} & \(\log \left(\det (X)\right)\) & \(X \in \mathbf{S}^n_+\) & $\pm$ & concave & none\\
      \hline
      \code{log\_sum\_exp(X)} & \(\log \left(\sum_{ij}e^{X_{ij}}\right)\) & \(X \in\mathbf{R}^{m \times n}\) & $\pm$ & convex & $\nearrow$\\
      \hline
      \code{matrix\_frac(x, P)} & \(x^\top P^{-1} x\) & \(x \in \mathbf{R}^n\), \newline \(P \in\mathbf{S}^n_{++}\) & $+$ & convex & none\\
      \hline
      \code{max\_entries(X)} & \(\max_{ij}\left\{ X_{ij}\right\}\) & \(X \in\mathbf{R}^{m \times n}\) & same as $X$ & convex & $\nearrow$\\
      \hline
      \code{min\_entries(X)} & \(\min_{ij}\left\{ X_{ij}\right\}\) & \(X \in\mathbf{R}^{m \times n}\) & same as $X$ & concave & $\nearrow$\\
      \hline
      \code{mixed\_norm(X, p, q)} & \(\left(\sum_k\left(\sum_l\lvert x_{k,l}\rvert^p\right)^{q/p}\right)^{1/q}\) & \(X \in\mathbf{R}^{n \times n}\) & $+$ & convex & none\\
      \hline
      \code{cvxr\_norm(x)} \newline \code{cvxr\_norm(x, 2)} &
                                                     \(\sqrt{\sum_{i}x_{i}^2
                                                     }\) & \(X
                                                           \in\mathbf{R}^{n}\)
                                  & $+$ & convex & $\nearrow$ for
                                                   \(x_{i} \geq 0\), \newline $\swarrow$  for \(x_{i} \leq 0\)\\
      \hline
      \code{cvxr\_norm(X, "fro")} & \(\sqrt{\sum_{ij}X_{ij}^2 }\) &
                                                                    \(X
                                                                    \in\mathbf{R}^{m
                                                                    \times
                                                                    n}\)
                                  & $+$ & convex & $\nearrow$ for \(X_{ij} \geq 0\), $\swarrow$  for \(X_{ij} \leq 0\)\\
      \hline
      \code{cvxr\_norm(X, 1)} & \(\sum_{ij}\lvert X_{ij} \rvert\) &
                                                                    \(X
                                                                    \in\mathbf{R}^{m
                                                                    \times
                                                                    n}\)
                                  & $+$ & convex & $\nearrow$ for
                                                   \(X_{ij} \geq 0\), $\swarrow$  for \(X_{ij} \leq 0\)\\
      \hline
      \code{cvxr\_norm(X, "inf")} & \(\max_{ij} \{\lvert X_{ij}
                                    \rvert\}\) & \(X \in\mathbf{R}^{m
                                                 \times n}\) & $+$ &
                                                                     convex
                                                     & $\nearrow$ for
                                                       \(X_{ij} \geq
                                                       0\), $\swarrow$  for \(X_{ij} \leq 0\)\\
      \hline
      \code{cvxr\_norm(X, "nuc")} & \(\mathrm{tr}\left(\left(X^\top X\right)^{1/2}\right)\) & \(X \in\mathbf{R}^{m \times n}\) & $+$ & convex & none\\
      \hline
      \code{cvxr\_norm(X)} \newline \code{cvxr\_norm(X, 2)} & \(\sqrt{\lambda_{\max}\left(X^\top X\right)}\) & \(X \in\mathbf{R}^{m \times n}\) & $+$ & convex & none\\
      \hline
    \end{tabular}
  \caption{Scalar functions.}
  \label{tab:scalar1}
\end{sidewaystable}

\begin{sidewaystable}
  \small
  \centering
  \begin{tabular}{p{1.85in}p{1.6in}p{.9in}p{.8in}p{.6in}p{1in}}      
    \hline
      Function & Meaning & Domain & Sign & Curvature & Monotonicity\\
      \hline
      \code{p\_norm(X, p)} \newline \(p \geq 1\) or \(p = \infty\) & \(\|X\|_p
                                                            =
                                                            \left(\sum_{ij}
                                                            |X_{ij}|^p
                                                            \right)^{1/p}\)
                         & \(X \in \mathbf{R}^{m \times n}\) & $+$ &
                                                                     convex
                                                     & $\nearrow$ for
                                                       \(X_{ij} \geq
                                                       0\), $\swarrow$  for \(X_{ij} \leq 0\)\\
      \hline
      \code{p\_norm(X, p)} \newline \(p < 1\), \(p \neq 0\) & \(\|X\|_p = \left(\sum_{ij} X_{ij}^p \right)^{1/p}\) & \(X \in \mathbf{R}^{m \times n}_+\) & $+$ & concave & $\nearrow$\\
      \hline
      \code{quad\_form(x, P)} \newline constant \(P \in \mathbf{S}^n_+\) &
                                                                  \(x^\top
                                                                  P
                                                                  x\)
                         & \(x \in \mathbf{R}^n\) & $+$ & convex &
                                                                   $\nearrow$ for \(x_i \geq 0\), \newline $\swarrow$  for \(x_i \leq 0\)\\
      \hline
      \code{quad\_form(x, P)} \newline constant \(P \in \mathbf{S}^n_-\) &
                                                                  \(x^\top
                                                                  P
                                                                  x\)
                         & \(x \in \mathbf{R}^n\) & $-$ & concave &
                                                                    $\nearrow$ for \(x_i \geq 0\), \newline $\swarrow$ for \(x_i \leq 0\)\\
      \hline
      \code{quad\_form(c, X)} \newline constant \(c \in \mathbf{R}^n\) & \(c^\top X c\) & \(X \in\mathbf{R}^{n \times n}\) & depends on $c$, $X$ & affine & depends on $c$\\
      \hline
      \code{quad\_over\_lin(X, y)} & \(\left(\sum_{ij}X_{ij}^2\right)/y\) &
                                                                     \(x
                                                                            \in
                                                                            \mathbf{R}^n\),
                                                                            \(y
                                                                            >
                                                                            0\)
                                  & $+$ & convex & $\nearrow$ for
                                                   \(X_{ij} \geq 0\),
                                                   $\swarrow$ for \(X_{ij} \leq 0\), $\swarrow$  in \(y\) \\
      \hline
      \code{sum\_entries(X)} & \(\sum_{ij}X_{ij}\) & \(X \in\mathbf{R}^{m \times n}\) & same as $X$ & affine & $\nearrow$\\
      \hline
      \code{sum\_largest(X, k)} \newline \(k = 1,2,\ldots\) & \(\text{sum of } k\text{ largest }X_{ij}\) & \(X \in\mathbf{R}^{m \times n}\) & same as $X$ & convex & $\nearrow$\\
      \hline
      \code{sum\_smallest(X, k)} \newline \(k = 1,2,\ldots\) & \(\text{sum of } k\text{ smallest }X_{ij}\) & \(X \in\mathbf{R}^{m \times n}\) & same as $X$ & concave & $\nearrow$\\
      \hline
      \code{sum\_squares(X)} & \(\sum_{ij}X_{ij}^2\) & \(X \in\mathbf{R}^{m
                                                \times n}\) & $+$ &
                                                                    convex
                                                     & $\nearrow$ for
                                                       \(X_{ij} \geq
                                                       0\), $\swarrow$  for \(X_{ij} \leq 0\)\\
      \hline
      \code{matrix\_trace(X)} & \(\mathrm{tr}\left(X \right)\) & \(X \in\mathbf{R}^{n \times n}\) & same as $X$ & affine & $\nearrow$\\
      \hline
      \code{tv(x)} & \(\sum_{i}|x_{i+1} - x_i|\) & \(x \in \mathbf{R}^n\) & $+$ & convex & none\\
      \hline
      \code{tv(X)}
               & \begin{minipage}{1.6in}\vskip5pt\(\sum_{ij}\left\|
                   \left[\begin{matrix} X_{i+1,j} - X_{ij} \\
                       X_{i,j+1} -X_{ij} \end{matrix}\right]
                 \right\|_2\) \\[5pt] \end{minipage}& \(X \in \mathbf{R}^{m \times n}\) & $+$ & convex & none\\
      \hline
      \code{tv(X1,\ldots,Xk)}
               & \begin{minipage}{1.6in}\vskip5pt\(\sum_{ij}\left\|
                   \left[\begin{matrix} X_{i+1,j}^{(1)} - X_{ij}^{(1)}
                       \\ X_{i,j+1}^{(1)} -X_{ij}^{(1)} \\ \vdots \\
                       X_{i+1,j}^{(k)} - X_{ij}^{(k)} \\
                       X_{i,j+1}^{(k)}
                       -X_{ij}^{(k)} \end{matrix}\right] \right\|_2\)
                 \\[5pt] \end{minipage}& \(X^{(i)} \in\mathbf{R}^{m
                                           \times n}\) & $+$ & convex
                                         & none\\
    \hline
  \end{tabular}
  \caption{More scalar functions.}
  \label{tab:scalar2}
\end{sidewaystable}

\section{Expressions and functions}

\cvxr{} uses the function information in this section and the DCP
tools to assign expressions a sign and curvature. In what follows,
the domain $\symm^n$ refers to the set of symmetric matrices,
with $\symm_+^n$ and $\symm_-^n$ referring to the set of
positive semidefinite and negative semidefinite matrices,
respectively.

\subsection{Operators}

The infix operators \code{+}, \code{-}, \code{*}, \code{\%*\%},
\code{/} are treated as functions. Both \code{+} and \code{-} are
affine functions. In \cvxr{}, \code{*} and \code{/} are affine because \code{expr1 * expr2} and \code{expr1 \%*\% expr2} are allowed only when one of the expressions is constant and \code{expr1 / expr2}
is allowed only when \code{expr2} is a scalar constant.

The transpose of any expression can be obtained using
\code{t(expr)}. Transpose is an affine function. The construct 
\code{expr\^{}p} is equivalent to the function \code{power(expr, p)}.

\subsection{Indexing and slicing}

All non-scalar expressions can be indexed using
\code{expr[i, j]}. Indexing is an affine function. The syntax
\code{expr[i]} can be used as a shorthand for \code{expr[i, 1]} when
\code{expr} is a column vector. Similarly, \code{expr[i]} is shorthand
for \code{expr[1, i]} when \code{expr} is a row vector.

Non-scalar expressions can also be sliced using the standard
\proglang{R} slicing syntax. For example, \code{expr[i:j, r]} selects
rows \code{i} through \code{j} of column \code{r} and returns a
vector.

\cvxr{} supports advanced indexing using lists of indices or boolean
arrays. The semantics are the same as in \proglang{R}. Any time
\proglang{R} might return a numeric vector, \cvxr{} returns a column
vector.

\begin{sidewaystable}
  \small
  \centering
  \begin{tabular}{p{1.85in}p{1.6in}p{.9in}p{.8in}p{.6in}p{1in}}      
\hline
Function & Meaning & Domain & Sign & Curvature & Monotonicity\\
\hline
\code{abs(x)} & \(\lvert x \rvert\) & \(x \in \mathbf{R}\) & $+$ &
                                                                   convex
                                               & $\nearrow$ for \(x
                                                 \geq 0\), $\swarrow$  for \(x \leq 0\)\\
\hline
\code{entr(x)} & \(-x \log (x)\) & \(x > 0\) & $\pm$ & concave & none\\
\hline
\code{exp(x)} & \(e^x\) & \(x \in \mathbf{R}\) & $+$ & convex & $\nearrow$\\
\hline
\code{huber(x, M = 1)} \newline \(M \geq 0\) & \(\begin{cases}x^2 &|x| \leq M \\
  2M|x| - M^2&|x| > M\end{cases}\) & \(x \in \mathbf{R}\) & $+$ &
                                                                  convex & $\nearrow$ for \(x \geq 0\), $\swarrow$   for \(x \leq 0\)\\
\hline
\code{inv\_pos(x)} & \(1/x\) & \(x > 0\) & $+$ & convex & $\swarrow$\\
\hline
\code{kl\_div(x, y)} & \(x \log(x/y) - x + y\) & \(x > 0\), \(y > 0\) & $+$ & convex & none\\
\hline
\code{log(x)} & \(\log(x)\) & \(x > 0\) & $\pm$ & concave & $\nearrow$\\
\hline
\code{log1p(x)} & \(\log(x+1)\) & \(x > -1\) & same as $x$ & concave & $\nearrow$\\
\hline
\code{logistic(x)} & \(\log(1 + e^{x})\) & \(x \in \mathbf{R}\) & $+$ & convex & $\nearrow$\\
\hline
\code{max\_elemwise(x1,..., xk)} & \(\max \left\{x_1, \ldots , x_k\right\}\) & \(x_i \in \mathbf{R}\) & \(\max(\mathrm{sign}(x_1))\) & convex & $\nearrow$\\
\hline
\code{min\_elemwise(x1,..., xk)} & \(\min \left\{x_1, \ldots , x_k\right\}\) & \(x_i \in \mathbf{R}\) & \(\min(\mathrm{sign}(x_1))\) & concave & $\nearrow$\\
\hline
\code{multiply(c, x)} \newline \(c \in \mathbf{R}\) & \(c \times x\) & \(x \in\mathbf{R}\) & \(\mathrm{sign}(cx)\) & affine & depends on c\\
\hline
\code{neg(x)} & \(\max \left\{-x, 0 \right\}\) & \(x \in \mathbf{R}\) & $+$ & convex & $\swarrow$\\
\hline
\code{pos(x)} & \(\max \left\{x, 0 \right\}\) & \(x \in \mathbf{R}\) & $+$ & convex & $\nearrow$\\
\hline
\code{power(x, 0)} & \(1\) & \(x \in \mathbf{R}\) & $+$ & constant & \\
\hline
\code{power(x, 1)} & \(x\) & \(x \in \mathbf{R}\) & same as $x$ & affine & $\nearrow$\\
\hline
\code{power(x, p)} \newline \(p = 2, 4, 8, \ldots\) & \(x^p\) & \(x \in
                                                       \mathbf{R}\) &
                                                                      $+$ & convex & $\nearrow$ for \(x \geq 0\), $\swarrow$ for \(x \leq 0\)\\
\hline
\code{power(x, p)} \newline \(p < 0\) & \(x^p\) & \(x > 0\) & $+$ & convex & $\swarrow$\\
\hline
\code{power(x, p)} \newline \(0 < p < 1\) & \(x^p\) & \(x \geq 0\) & $+$ & concave & $\nearrow$\\
\hline
\code{power(x, p)} \newline \(p > 1,\ p \neq 2, 4, 8, \ldots\) & \(x^p\) & \(x \geq 0\) & $+$ & convex & $\nearrow$\\
\hline
\code{scalene(x, alpha, beta)} \newline \(\text{alpha} \geq 0\), \(\text{beta}
    \geq 0\) & \(\alpha\mathrm{pos}(x)+ \beta\mathrm{neg}(x)\) & \(x
                                                                 \in
                                                                 \mathbf{R}\)
                            & $+$ & convex & $\nearrow$ for \(x \geq
                                             0\), $\swarrow$ for \(x \leq 0\)\\
\hline
\code{sqrt(x)} & \(\sqrt x\) & \(x \geq 0\) & $+$ & concave & $\nearrow$\\
\hline
\code{square(x)} & \(x^2\) & \(x \in \mathbf{R}\) & $+$ & convex & $\nearrow$ for \(x \geq 0\), $\swarrow$ for \(x \leq 0\)\\
\hline
\end{tabular}
  \caption{Elementwise functions.}
  \label{tab:elemwise}
\end{sidewaystable}

\begin{sidewaystable}
  \small
  \centering
  \begin{tabular}{p{1.85in}p{1.6in}p{.9in}p{.8in}p{.6in}p{1in}}      
    \hline
    Function & Meaning & Domain & Sign & Curvature & Monotonicity\\
    \hline
    \code{bmat([[X11,..., X1q], ..., [Xp1,..., Xpq]])} & \(\left[\begin{matrix} X^{(1,1)} & \cdots & X^{(1,q)} \\ \vdots & & \vdots \\ X^{(p,1)} & \cdots & X^{(p,q)} \end{matrix}\right]\) & \(X^{(i,j)} \in\mathbf{R}^{m_i \times n_j}\) & \(\mathrm{sign}\left(\sum_{ij} X^{(i,j)}_{11}\right)\) & affine & $\nearrow$\\
    \hline
    \code{conv(c, x)} \newline \(c\in\mathbf{R}^m\) & \(c*x\) & \(x\in \mathbf{R}^n\) & \(\mathrm{sign}\left(c_{1}x_{1}\right)\) & affine & depends on $c$\\
    \hline
    \code{cumsum\_axis(X, axis = 1)} & cumulative sum along given axis & \(X \in \mathbf{R}^{m \times n}\) & same as X & affine & $\nearrow$\\
    \hline
    \code{diag(x}) & \(\left[\begin{matrix}x_1 & & \\& \ddots & \\& & x_n\end{matrix}\right]\) & \(x \in\mathbf{R}^{n}\) & same as x & affine & $\nearrow$\\
    \hline
    \code{diag(X)} & \(\left[\begin{matrix}X_{11} \\\vdots \\X_{nn}\end{matrix}\right]\) & \(X \in\mathbf{R}^{n \times n}\) & same as X & affine & $\nearrow$\\
    \hline
    \code{diff(X, k = 1, axis = 1)} \newline \(k = 0,1,2,\ldots\) & \(k\)th order differences (argument \(k\) is actually named \emph{differences} and \emph{lag} can also be used) along given axis & \(X \in\mathbf{R}^{m \times n}\) & same as X & affine & $\nearrow$\\
    \hline
    \code{hstack(X1,..., Xk)} & \(\left[\begin{matrix}X^{(1)} \cdots X^{(k)}\end{matrix}\right]\) & \(X^{(i)} \in\mathbf{R}^{m \times n_i}\) & \(\mathrm{sign}\left(\sum_i X^{(i)}_{11}\right)\) & affine & $\nearrow$\\
    \hline
    \code{kronecker(C, X)} \newline \(C\in\mathbf{R}^{p \times q}\) & \(\left[\begin{matrix}C_{11}X & \cdots & C_{1q}X \\ \vdots & & \vdots \\ C_{p1}X & \cdots & C_{pq}X \end{matrix}\right]\) & \(X \in\mathbf{R}^{m \times n}\) & \(\mathrm{sign}\left(C_{11}X_{11}\right)\) & affine & depends on $C$\\
    \hline
    \code{reshape\_expr(X, c(m', n'))} & \(X' \in\mathbf{R}^{m' \times n'}\) & \(X \in\mathbf{R}^{m \times n}\) \(m'n' = mn\) & same as X & affine & $\nearrow$\\
    \hline
    \code{vec(X)} & \(x' \in\mathbf{R}^{mn}\) & \(X \in\mathbf{R}^{m \times n}\) & same as X & affine & $\nearrow$\\
    \hline
    \code{vstack(X1,..., Xk)} & \(\left[\begin{matrix}X^{(1)} \\ \vdots \\X^{(k)}\end{matrix}\right]\) & \(X^{(i)} \in\mathbf{R}^{m_i \times n}\) & \(\mathrm{sign}\left(\sum_i X^{(i)}_{11}\right)\) & affine & $\nearrow$\\
    \hline
\end{tabular}
  \caption{Vector and matrix functions.}
  \label{tab:vecfns}
\end{sidewaystable}

\subsection{Scalar functions}

\cvxr{} provides the scalar functions displayed in
Tables~\ref{tab:scalar1} and~\ref{tab:scalar2}, which take in one or
more scalars, vectors, or matrices as arguments and return a scalar.

For a vector expression \code{x}, \code{cvxr_norm(x)} and
\code{cvxr_norm(x, 2)} give the Euclidean norm. For a matrix
expression \code{X}, however, \code{cvxr_norm(X)} and
\code{cvxr_norm(X, 2)} give the spectral norm.  The function
\code{cvxr_norm(X, "fro")} gives the Frobenius norm and
\code{cvxr_norm(X, "nuc")} the nuclear norm. The nuclear norm can also
be defined as the sum of the singular values of \code{X}.

The functions \code{max_entries} and \code{min_entries} give the
largest and smallest entry, respectively, in a single
expression. These functions should not be confused with
\code{max_elemwise} and \code{min_elemwise} (see Section~\ref{appendix:elementwise}). The
functions \code{max_elemwise} and \code{min_elemwise} return the maximum
or minimum of a list of scalar expressions.

The function \code{sum_entries} sums all the entries in a single
expression. The built-in \proglang{R} \code{sum} should be used to add
together a list of expressions. For example, the following code sums
three expressions.
\begin{CodeChunk}
\begin{CodeInput}
R> sum(expr1, expr2, expr3)
\end{CodeInput}
\end{CodeChunk}
Some functions such as \code{sum_entries}, \code{cvxr_norm},
\code{max_entries}, and \code{min_entries} can be applied along an
axis. Given an $m$ by $n$ expression \code{expr}, the line
\code{func(expr, axis = 1)} applies \code{func} to each row, returning
an $m$ by 1 expression. The line \code{func(expr, axis = 2)} applies
\code{func} to each column, returning a 1 by $n$ expression. For example, the
following code sums along the columns and rows of a matrix variable:
\begin{CodeChunk}
\begin{CodeInput}
R> X <- Variable(5, 4)
R> row_sums <- sum_entries(X, axis = 1)   # Has size (5, 1)
R> col_sums <- sum_entries(X, axis = 2)   # Has size (1, 4)
\end{CodeInput}
\end{CodeChunk}
\cvxr{} ensures the implementation aligns with the \code{base::apply}
function. The default in most cases is \code{axis = NA}, which treats
an input matrix as one long vector, basically the same as \code{base::apply}
with \code{MARGIN = c(1, 2)}. The exception is \code{cumsum_axis} (see
Table~\ref{tab:elemwise}), which cannot take \code{axis = NA} and will throw an error.

\subsection{Elementwise functions}\label{appendix:elementwise}

These functions operate on each element of their arguments and are
displayed in Table~\ref{tab:elemwise}. For
example, if \code{X} is a 5 by 4 matrix variable, then \code{abs(X)} is
a 5 by 4 matrix expression. Also, \code{abs(X)[1, 2]} is equivalent to
\code{abs(X[1, 2])}.

Elementwise functions that take multiple arguments, e.g.,
\code{max_elemwise} and \code{multiply}, operate on the
corresponding elements of each argument. For instance, if \code{X} and
\code{Y} are both 3 by 3 matrix variables, then \code{max\_elemwise(X,
  Y)} is a 3 by 3 matrix expression, where \code{max_elemwise(X, Y)[2,
  1]} is equivalent to \code{max\_elemwise(X[2, 1], Y[2, 1])}. Thus
all arguments must have the same dimensions or be scalars, which are
promoted appropriately.

\subsection{Vector and matrix functions}

These functions, shown in Table~\ref{tab:vecfns}, take one or more
scalars, vectors, or matrices as arguments and return a vector or
matrix.

The input to \code{bmat} is a list of lists of \cvxr{} expressions. It
constructs a block matrix. The elements of each inner list are stacked
horizontally, and then the resulting block matrices are stacked
vertically.

The output of \code{vec(X)} is the matrix \code{X} flattened in
column-major order into a vector.

The output of \code{reshape_expr(X, c(m1, n1))} is the matrix \code{X} 
cast into an \code{m1} by \code{n1} matrix. The entries are taken from
\code{X} in column-major order and stored in the output in column-major order.

\end{document}